\documentclass{aa}
\usepackage{txfonts}
\usepackage{graphicx}
\usepackage{psfig,longtable,lscape}
\usepackage{epsfig}
\usepackage[ ]{natbib} 
\usepackage{multirow}
\def\underddot#1{\mathord{\vtop to0pt{\ialign{##\crcr
$\hfil\displaystyle{#1}\hfil$\crcr\noalign{\kern1.5pt\nointerlineskip}
$\hfil{.\kern-0.7pt.}{}\kern1.5pt\hfil$\crcr}\vss}}}
\bibpunct{(}{)}{;}{a}{}{,} 

\def\underdddot#1{\mathord{\vtop to0pt{\ialign{##\crcr
$\hfil\displaystyle{#1}\hfil$\crcr\noalign{\kern1.5pt\nointerlineskip}
$\hfil{.\kern-0.7pt .\kern-0.7pt .}{}\kern1.5pt\hfil$\crcr}\vss}}}

\begin{document}

\title{X-ray spectroscopy  of the ADC source X1822-371 with Chandra and 
XMM-Newton}
  

\author{R. Iaria\inst{1}, T. Di Salvo\inst{1},  A. D'A\`\i\inst{1}, 
L. Burderi\inst{2},  
T. Mineo\inst{3},  A. Riggio\inst{2,4}, 
A. Papitto\inst{2,4}, N. R. Robba\inst{1}}


\offprints{R. Iaria, \email{iaria@fisica.unipa.it}}

\institute{Dipartimento di Scienze Fisiche ed Astronomiche,
Universit\`a di Palermo, via Archirafi 36 - 90123 Palermo, Italy
\and 
Dipartimento di Fisica, Universit\`a degli Studi di Cagliari, SP Monserrato-Sestu, KM 0.7, Monserrato, 09042 Italy
  \and
  INAF, Istituto di Astrofisica Spaziale e Fisica cosmica di Palermo, 
via U. La Malfa 153, I-90146 Palermo, Italy
  \and 
 INAF, Osservatorio Astronomico di Cagliari, Poggio dei Pini, Strada 54, 09012 Capoterra (CA), Italy
} 

\date{}

\abstract
{ The eclipsing low-mass X-ray binary X1822-371 is the prototype of
  the accretion disc corona (ADC) sources. Its inclination angle
  ($\simeq 82.5^{\circ}$) is high enough that flux from the neutron
  star is blocked by the edge-on accretion disc.  Because the
    neutron star's direct emission is hidden, its ADC emission is
    visible.   }
{ We analyse two Chandra observations and one XMM-Newton observation 
  to study the discrete features  in this source and their variation as a
  function of the orbital phase, deriving  constraints on the
  temperature, density, and location of the plasma responsible for emission 
  lines.}
{The HETGS and XMM/Epic-pn observed X1822-371 for 140 and 50 ks,
  respectively. We extracted an averaged spectrum and five spectra
  from five selected orbital-phase intervals that are 0.04-0.25,
  0.25-0.50, 0.50-0.75, 0.75-0.95, and, finally, 0.95-1.04; the
  orbital phase zero corresponds to the eclipse time.  All spectra
  cover the energy band between 0.35 and 12 keV.  }
{ We confirm the presence of local neutral matter that partially
  covers the X-ray emitting region; the equivalent hydrogen column is
  $5 \times 10^{22}$ cm$ ^{-2}$ and the covered fraction is about
  60--65\%. We identify emission lines from \ion{O}{vii},
  \ion{O}{viii}, \ion{Ne}{ix} \ion{Ne}{x}, \ion{Mg}{xi},
  \ion{Mg}{xii}, \ion{Si}{xiii}, \ion{Si}{xiv}, \ion{Fe}{xxv},
  \ion{Fe}{xxvi}, and a prominent fluorescence iron line associated
  with a blending of \ion{Fe}{i}-\ion{Fe}{xv} resonant transitions.
  The transitions of He-like ions show that the intercombination
  dominates over the forbidden and resonance lines. The line fluxes
  are the highest during the orbital phases between 0.04 and 0.75.}
{We discuss the presence of an extended, optically thin corona with
  optical depth of about 0.01 that scatters the X-ray photons from the
  innermost region into the line of sight. The photoionised plasma
  producing the \ion{O}{viii}, \ion{Ne}{ix}, \ion{Ne}{x},
  \ion{Mg}{xi}, \ion{Mg}{xii}, \ion{Si}{xiii}, and \ion{Si}{xiv} lines
  is placed in the bulge at the outer radius of the disc distant from
  the central source of $6 \times 10^{10}$ cm. The \ion{O}{vii} and
  the fluorescence iron line are probably produced in the photoionised
  surface of the disc at inner radii.  Finally, we suggest that the
  observed local neutral matter is the matter transferred by the
  companion star that was expelled from the system by the X-ray
  radiation pressure, which in turn originated in the accretion
  process onto the neutron star. (Abridged)}
\keywords{line: identification -- line: formation -- stars: individual
(X1822-371)  --- X-rays: binaries  --- X-rays: general}
\authorrunning {R.\ Iaria et al.}
\titlerunning {X-ray spectroscopy of the ADC source X1822-371}

\maketitle

\section{Introduction}

X1822-371 is a compact binary system with a period of 5.57 hr.  The
light curve of the source shows an almost sinusoidal modulation and
partial eclipse. The partial eclipse indicates a high inclination angle of the
system. Initially, \cite{Mason82} derived an inclination angle between
76$^\circ$ and 84$^\circ$ fitting the light curve of X1822-371 in the
infrared band.  The modulation of the light curve is generally
explained with the presence of a geometrically thick disc whose height
varies depending on the azimuthal angle and occults part of the X-ray
emission \citep{White_holt_1982,Mason82,
  Hellier_mason1989,Bayless2010}.

Analysing RXTE data of X1822-371, \cite{Jonker_2001} detected a
coherent pulsation at 0.59 s, and derived that the orbit of the system
is almost circular with an eccentricity less than 0.03, a mass
function of $(2.03 \pm 0.03) \times 10^{-2}$ M$_\odot$, and a pulse
period derivative of $(-2.85 \pm 0.04) \times 10^{-12}$ s s$^{-1}$,
indicating that the neutron star is spinning up.  The distance to
X1822-371 is 2.5 kpc \citep{Mason82} and the observed luminosity is $
10^{36}$ erg s$^{-1}$ \citep[see e.g.][]{Mason82, Iaria2001_1822}.
However, \cite{Jonker_2001}, using the relation between luminosity and
spin-up rate, found that if X1822-371 is at a luminosity of $ 10^{36}$
erg s$^{-1}$ the neutron star magnetic field $B$ should have an
unlikely strength of $\sim 8 \times 10^{16}$ G, while for a luminosity
of $ 10^{38}$ erg s$^{-1}$ $B \sim 8 \times 10^{10}$ G.
\cite{Jonker_2003} constrained the inclination angle of the source
 at $82.5^{\circ} \pm 1.5^{\circ}$ using ultraviolet
 and visual data of X1822-371.
 This value implies that the modulation in the light
curve of X1822-371 is not caused by dips because these can be observed
for inclination angles between 60$^{\circ}$ and 80$^{\circ}$
\citep{Frank}.

\cite{Burderi_2010} analysed the eclipse arrival times of X1822-371
using data from RXTE, XMM-Newton, and Chandra observations, spanning
the years from 1996 to 2008. Combining these eclipse arrival time
measurements with those already available \citep[covering the period
from 1977 to 1996; see][]{parmar2000}, the authors were able to
tightly constrain the orbital period derivative of the binary system
to $\dot{P}_{\rm orb} = 1.50(7) \times 10^{-10}$ s/s, that is three
orders of magnitude larger than  expected from conservative
mass transfer driven by magnetic braking and gravitational radiation.
\cite{Burderi_2010} concluded that the mass transfer rate from the
companion star is between 3.5 and 7.5 times the Eddington limit (i. e. $
1.8 \times 10^{-8}$ M$_\odot$ yr$^{-1}$ for a neutron star of 1.4 M$_\odot$),
suggesting that the mass transfer has to be highly non-conservative,
with the neutron star accreting at the Eddington limit and the rest of
the transferred mass expelled from the system.

\cite{Bayless2010}, studying the optical and UV data of X1822-371,
derived the new optical ephemeris for the source, finding a rapid
change of the orbital period with $P/\dot{P} = (3.0 \pm 0.3) \times
10^6 $ yr, similar to the value of the recent X-ray ephemeris of 
X1822-371 obtained by \cite{Burderi_2010}.  \cite{Bayless2010} showed
that the accretion rate onto the neutron star should be $6.4 \times
10^{-8}$ M$_\odot$ yr$^{-1}$ in a conservative mass transfer,
suggesting again a highly non-conservative mass transfer.  
 Finally they fitted the optical light curve of X1822-371,
concluding that the disc has a vertically extended component that is
optically thick at the optical wavelengths with a height of 0.5 times
the accretion disc radius. The authors identified this component as
the base of the disc wind.

Recently, \cite{Iaria_2011} extended the work of \cite{Burderi_2010},
including a larger sample of data, and combined  the optical/UV
data from \cite{Bayless2010} to obtain an updated ephemeris of
X1822-371.  \cite{Iaria_2011} obtained $\dot{P}_{\rm orb} = 1.59(9)
\times 10^{-10}$ s/s from the combined X-ray/optical/UV data set,
consistent with the previous measurements.  A similar value of
  the orbital period derivative ($\dot{P}_{\rm orb} = 1.3(3) \times
  10^{-10}$ s/s) was independently obtained by \cite{jain2010}
  by analysing data in the X-ray band.

Because of the high inclination angle of the system and the partial
eclipse, it has been argued that we do not observe the direct X-ray
emission produced in the inner region of the system, but the emission
coming from an extended corona above the disc, the so-called accretion
disc corona \citep[ADC,][]{White_holt_1982}. The properties of the ADC
are still debated. In a scenario with an optically thin ADC we should
expect that only a small fraction of the emission produced in the
inner region is scattered by the ADC into the line of sight \citep[see
e.g.][]{McClintock1982,Hellier_mason1989}. On the other hand, an
optically thick corona would imply an isotropic re-emission of the
primary flux \citep[][]{White_holt_1982}.

 \begin{figure*}[ht]
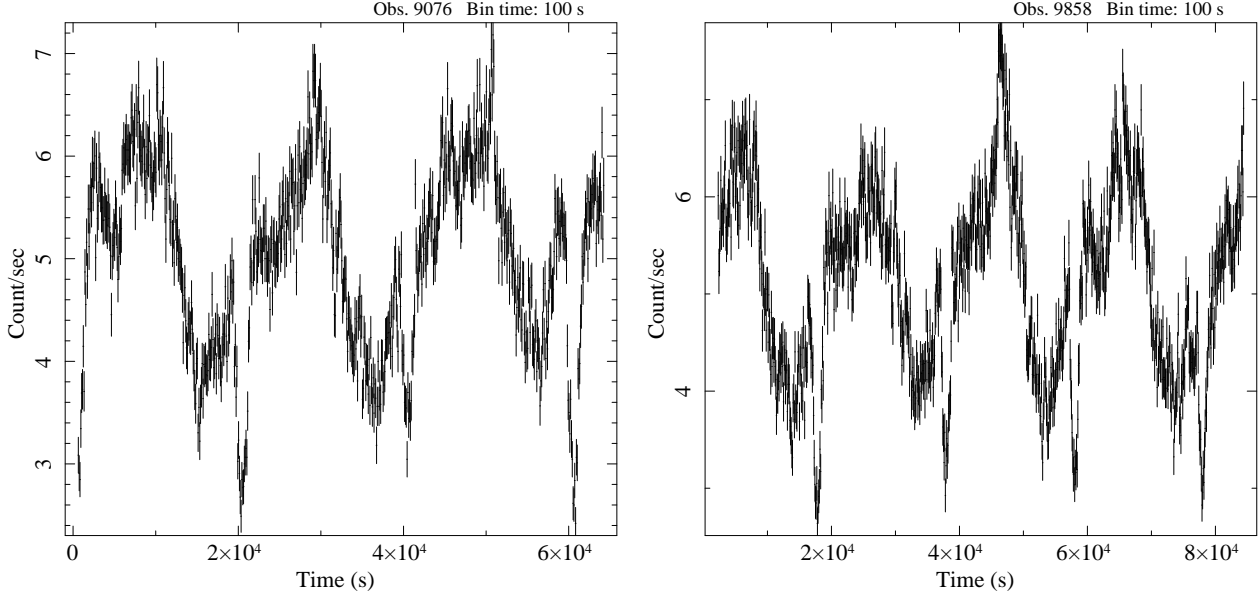

\includegraphics[height=8cm,angle=0]{fig1a.ps}
\includegraphics[height=8cm,angle=0]{fig1b.ps}
\caption{First-order MEG+HEG light curves of X1822-371.
The light curves correspond to Obs. ID. 9076 (left panel) and  
9858 (right panel), respectively.  The bin time is 100 s 
for both  light curves.}
\label{Fig1}
\end{figure*}

The X-ray spectrum of X1822-371 has been widely debated.
\cite{Hellier_mason1989} fitted the 1-10 keV EXOSAT spectrum of
X1822-371 with a black body peaked at 2 keV plus a flat power-law
component and an iron emission line at 6.7 keV.  The authors inferred
that the area of the emitting black-body surface was smaller ($\sim
1/400$) than the neutron star surface. They proposed that the
black-body emission originates from the neutron star surface but the
high inclination angle of the system and the tall outer disc would
block part of the direct emission from the neutron star, and only a
small percentage of the radiation is visible to the observer because it
is scattered by an optically thin corona. The power law component was
associated with the emission from the inner accretion disc.

\cite{Heinz_nowak_2001} analysed simultaneous observations with {\it
  Rossi X-Ray Timing Explorer }(RXTE) and ASCA of X1822-371. They
showed that both the source spectrum and light curve can be well-fitted 
by two equivalent models, representing the scenario in which
X1822-371 has an optically thick corona or an optically thin corona,
respectively. In the first case, no soft thermal component from the
inner region contributes to the source spectrum, and the emission from
the optically thick corona is described by a cutoff power law
partially absorbed by a cold gas that is the atmosphere of the outer
disc. In the second case, the model consists of a black-body component
emitted from the central region, the neutron star surface and/or the
inner disc, plus a Comptonised component produced in the optically
thin corona and fitted with a cutoff power law in the spectrum.  Both
these models were able to describe the data adequately.

\cite{Iaria2001_1822} analysed a BeppoSAX observation of X1822-371,
finding that the continuum emission is well-fitted using a Comptonised
component with an electron temperature of 4.5 keV that is partially
covered by a local neutral absorber.  The authors suggested that the
Comptonised emission is caused by a diffuse emission from the ADC while
the partial covering is produced by a cold wall subtending an angle of
16$^{\circ}$ with respect to the equatorial plane and placed in the
external edge of the accretion disc.  \cite{parmar2000} fitted the
BeppoSAX data of X1822-371 adopting a continuum model composed of a
Comptonised component plus a black body.  The electron
temperature of the Comptonising region was $4.53 \pm 0.02$ keV, the
optical depth was $26.2 \pm 0.6$, the seed-photon temperature was
$0.15 \pm 0.02$ keV and the black-body temperature was $1.27 \pm 0.03$
keV.

\cite{Cottam2001}, analysing a Chandra observation of X1822-371 with
the High-Energy Transmission Grating Spectrometer (HETGS), identified several
emission lines produced in a photoionised plasma. The lines were
associated with highly ionised iron, silicon, magnesium, neon, and
oxygen, and with a prominent fluorescence iron line at 6.4 keV.  The
emission region of the photoionised lines was placed  at the inner
  part of the X-ray illuminated bulge that is at the point of impact
between the disc and the accretion stream from the companion star,
whereas the origin of the fluorescence iron line was located in an
extended region on the disc illuminated by the light scattered by the
corona.

 Recently \cite{Ji_2011}, analysing  Chandra data sets also used
  for this paper, measured the Doppler velocities of several lines.  They
  divided the observation into 40 intervals and moved a phase window
  with a width of 0.16 in phase across the 40 intervals, so
  that each interval had a phase shift of 0.025 from the previous, and
  therefore the intervals partially overlap in phase.  \cite{Ji_2011},
  measuring the Doppler velocity modulation of several ionised lines, inferred
  that the observed photo-ionised  emission lines originate from a
  confined region in the outer edge of the accretion disc near the hot
  spot.  The photo-ionised plasma where the lines originate was
  consistent with ionisation parameters $\xi >100$. \cite{Ji_2011},
  combining the disc size and reasonable assumptions for the plasma
  density (between $10^{12}$ and $10^{14}$ cm$^{-3}$), suggested that
  the illuminating disc luminosities are more than an order of magnitude
  higher than what is actually observed and concluded that the central
  emitting X-ray source is not directly observed.
 
We follow a different approach by obtaining information on the lines as a
function of the orbital phase from five independent intervals.  In
this paper we analyse two Chandra/HETGS observations and one
XMM-Newton observation of X1822-371 for a total observing time of 140
and 50 ks, respectively. We discuss the averaged spectrum and the
phase-resolved spectra and compare our results with past results
discussed in literature.
   
 \section{Observations}
 \subsection{Chandra observations}

\begin{figure*}
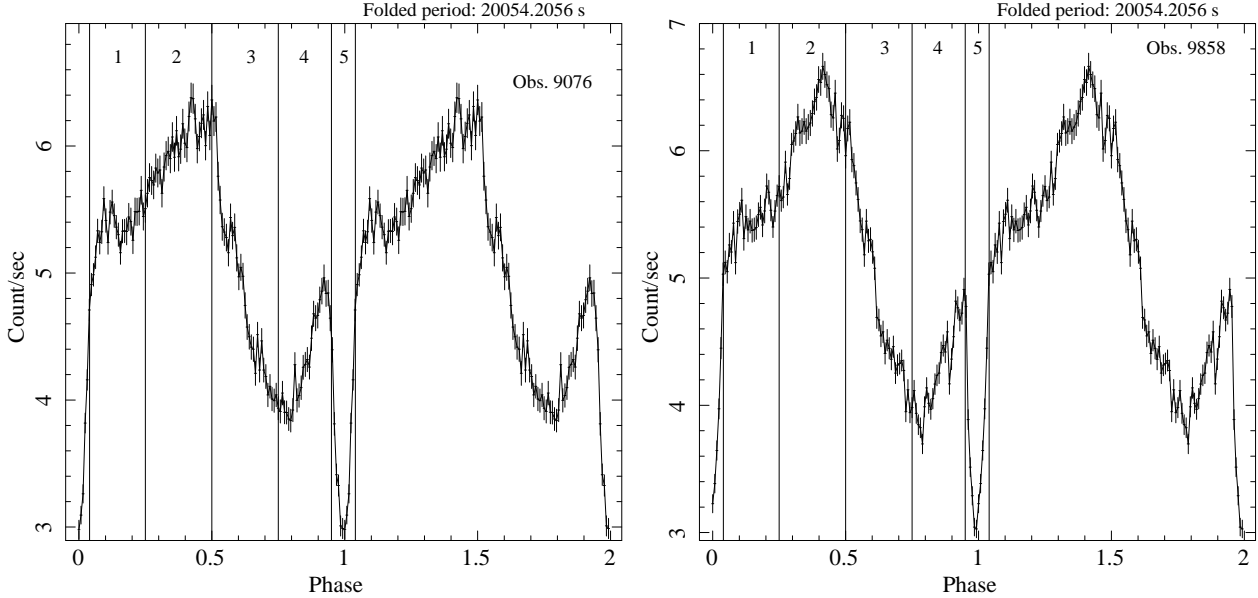

\includegraphics[height=8cm,angle=0]{fig2a.ps}
\includegraphics[height=8cm,angle=0]{fig2b.ps}
\caption{Folded light curves corresponding to  Obs. ID. 9076 (left
  panel) and Obs. ID. 9858 (right panel).  The data were folded
  adopting the X-ray ephemeris obtained by \cite{Iaria_2011} and using
  128 channels per period. The vertical lines indicate the
  phase intervals used for the phase-dependent spectral analysis (see
  section \ref{analysis}).}
\label{Fig2}
\end{figure*}

X1822-371 was observed with the Chandra observatory from 2008 May 20
22:53:00 to 2008 May 21 17:06:51 UT (Obs. ID. 9076) and from 2008 May
23 13:20:56 to 2008 May 24 12:41:54 UT (Obs. ID. 9858) using the HETGS
for a total observation time of 66 and 84 ks, respectively. Both
observations were performed in timed faint mode. The HETGS consists of
two types of transmission gratings, the medium-energy grating (MEG)
and the high-energy grating (HEG). The HETGS provides high-resolution
spectroscopy from 1.2 to 31 $\AA$ (0.4-10 keV) with a peak spectral
resolution of $\lambda / \Delta \lambda \sim 1000$ at 12 $\AA$ for the
HEG first order. The dispersed spectra were recorded with an array of
six charge-coupled devices (CCDs) that are part of the advanced CCD
imaging spectrometer-S \citep{Garmire2003}.  The current relative
accuracy of the overall wavelength calibration is 
0.05\%, leading to a worst-case uncertainty of 0.004 $\AA$ in the
first-order MEG and 0.006 $\AA$ in the first-order HEG.
The brightness of the source required additional efforts to mitigate
``photon pile-up'' effects of the zeroth-order events.  A 512 row
``subarray'' (with the first row = 1) was applied during the
observation, reducing the CCD frame time to 1.7 s. Pile-up distorts the
count spectrum because detected events overlap and their deposited
charges are collected into single, apparently more energetic, events.
We omitted the zeroth-order events from our spectral analysis and studied
only the grating first-order spectra, which were not affected by pile-up.

We  processed the event list using available software (FTOOLS
ver. 6.2 and CIAO ver. 4.1.2 packages) and computed aspect-corrected
exposure maps for each spectrum, which allowed us to take into account their 
effects on the effective area of the CCD spectrometer.  We barycentred the
events using the tool {\it axbary} in CIAO and show the 100 s bin time
light curves (merged HEG and MEG first-order events) corresponding to
 Obs. ID.  9076 and 9858 in Figure \ref{Fig1}. In Fig. \ref{Fig2}, we show
the folded the light curves adopting the recent X-ray ephemeris of
\cite{Iaria_2011} and using 128 channels per period. 
 The mean count rate of both observations is
5 cts/s out of the eclipse.

\subsection{XMM-Newton observation}
\begin{figure*}
\includegraphics[height=8cm,angle=0]{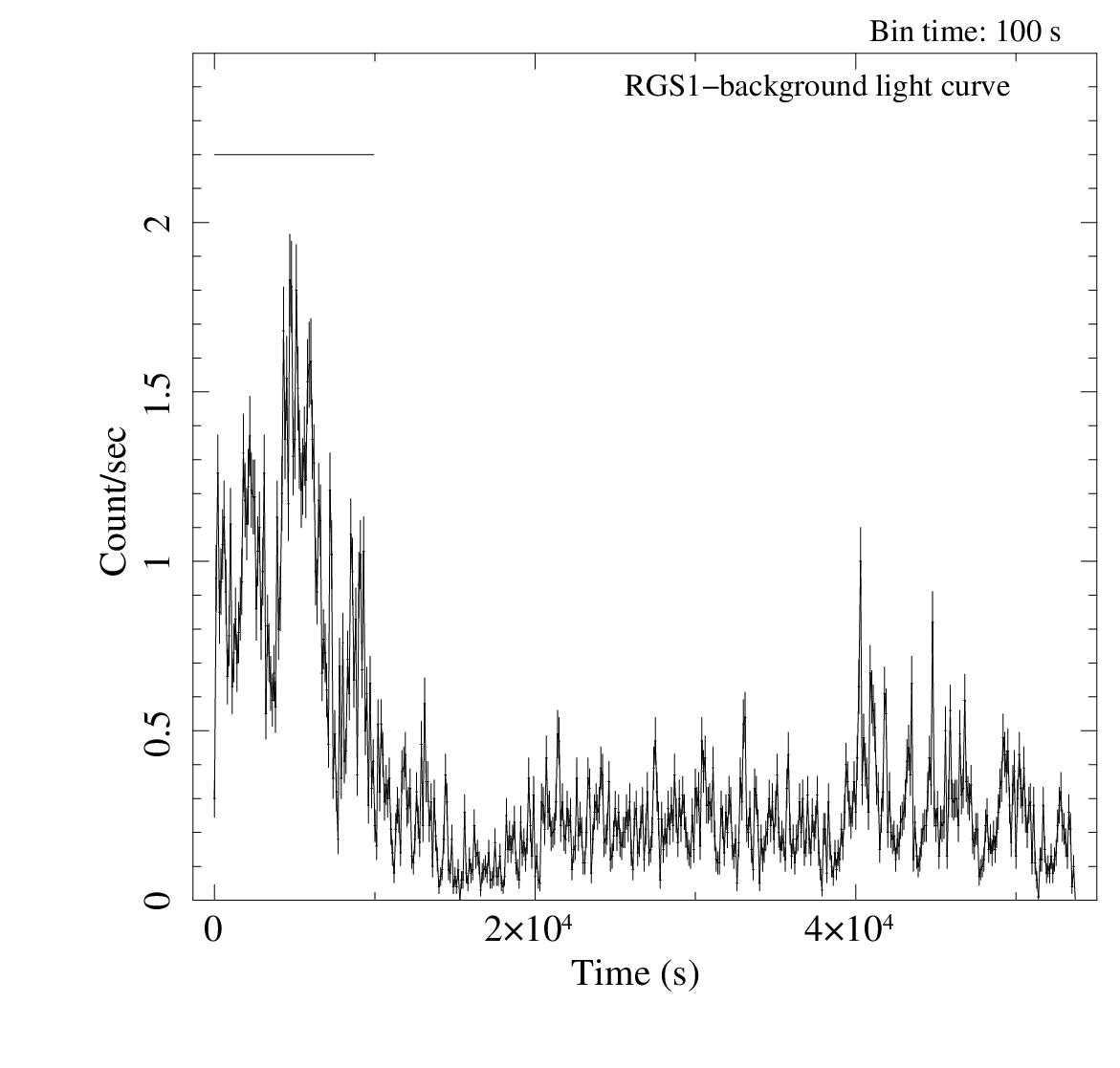}
\includegraphics[height=8cm,angle=0]{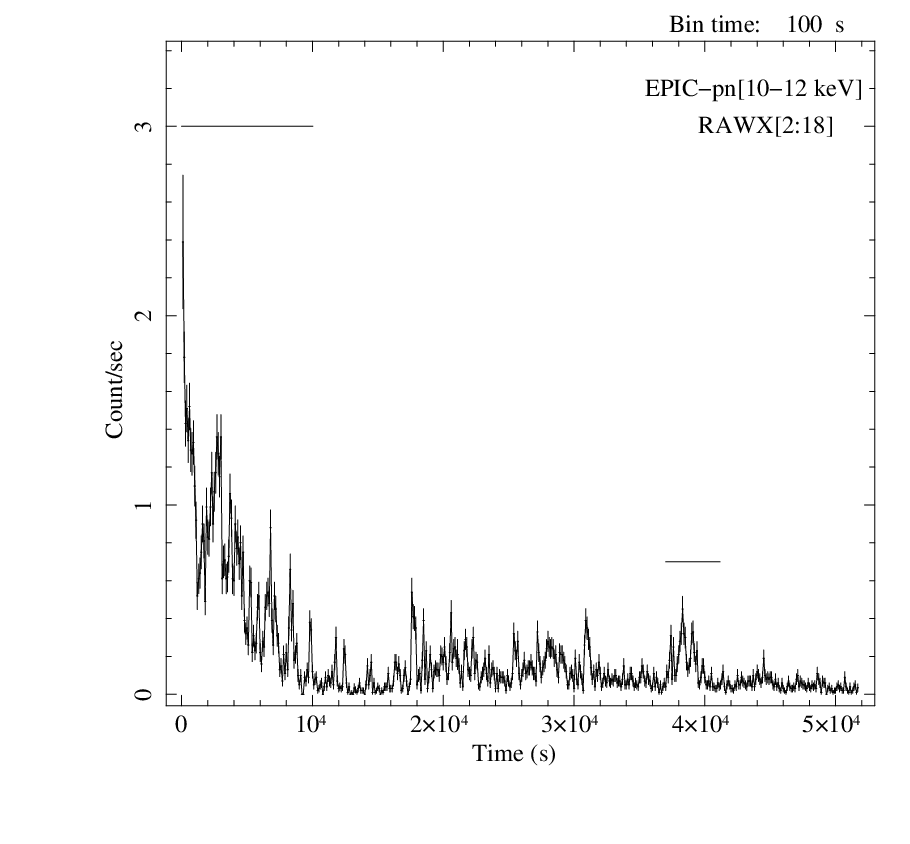}
\caption{RGS1 background light curve (left panel)
and  Epic-pn background light curve (right panel). The 
horizontal segments indicate the time interval excluded from our analysis.
The bin time is 100 s.}
\label{Fig_back_rgs1}
\end{figure*}

The XMM-Newton Observatory \citep{jansen01} includes three 1500 cm$^2$
X-ray telescopes each with an European Photon Imaging Camera (Epic,
0.1--15 keV) at the focus.  Two of the EPIC imaging spectrometers use
MOS CCDs \citep{turner01} and one uses pn CCDs \citep{struder01}.
Reflection grating spectrometers \citep[RGS, 0.35--2.5
keV,][]{denherder01} are located behind two of the telescopes.  The
region of sky containing X1822-371 was observed by XMM-Newton between
2001 March 07 13:12:48 UT and March 08 03:32:53 UT (Obs. ID. 0111230101)
for a duration of 53.8 ks. During the observation the MOS1 and MOS2 camera
were operated in fast uncompressed mode and small window mode,
respectively.  The Epic-pn camera was operated in timing mode with medium
filter during the observation. In this mode only one central CCD is
read out with a time resolution of 0.03 ms.  This provides a one
dimensional (4$\arcmin \!.$4 wide) image of the source with the second
spatial dimension being replaced by timing information.  The faster
CCD readout results in a much higher count rate capability of 800
cts/s before charge pile-up becomes a serious problem for point-like
sources. The Epic-pn count rate of the source was about 50 cts/s,
which avoids telemetry and pile-up problems. 

For our analysis we used only RGS and Epic-pn data since the Epic-pn
covers the same energy range as the MOS CCDs and has a larger
effective area.  We extracted the X-ray data products of RGS and
Epic-pn camera using the science analysis software (SAS) version 9.0.0
extracting only single and double events (patterns 0 to 4) from
Epic-pn data.  We selected source Epic-pn events from a $69.7\arcsec$
wide column (RAWX between 29 and 45) centred on the source position
(RAWX$=38$).

Using the SAS tools, we accumulated the RGS1 and Epic-pn background
light curves.  The RGS1 background light curve, plotted in Fig.
\ref{Fig_back_rgs1} (left panel), shows a flare (possibly due to solar
contamination) during the first 10 ks. We excluded this time interval
from our analysis because of its spurious origin.  The Epic-pn background
light curve was extracted from the columns between 2 and 18 and for
energies between 10 and 12 keV (see Fig.  \ref{Fig_back_rgs1}, right
panel). Together with the large flare in the first 10 ks, a smaller
flare is seen between 37 and 41 ks in the  Epic-pn light curve; we excluded
these two time intervals and barycentred the events using the tool
{\it barycen} in SAS to study the orbital-phase resolved spectrum.

\begin{figure}
\includegraphics[height=7.5cm,angle=0]{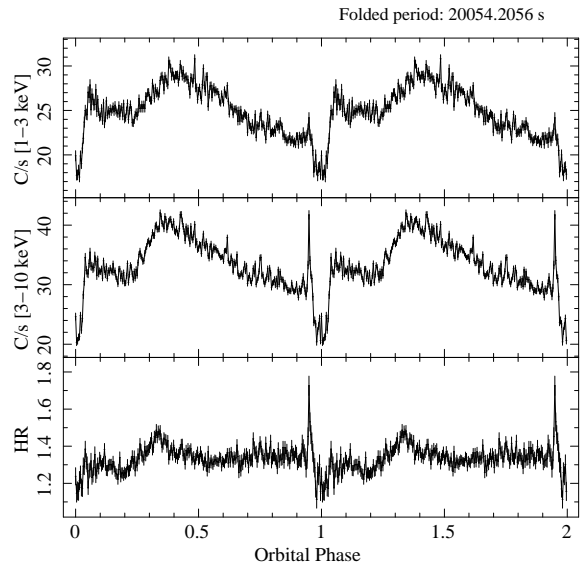}
\caption[]{ Upper (middle) panel: Epic-pn 1-3 keV (3-10 keV)
    folded light curves. Lower panel: Epic-pn hardness ratio for the
    two energy-selected light curves shown above. The data were folded
    adopting the recent X-ray ephemeris obtained by \cite{Iaria_2011}
    and using 256 channels per period.  }
\label{Fig_XMM_fold}
\end{figure}

Finally, we show in Fig. \ref{Fig_XMM_fold} the Epic-pn folded light
curves selected in the energy bands 1-3 keV and 3-10 keV, and the
folded hardness ratio (HR) obtained from the ratio of the 3-10
  keV folded light curve over the 1-3 keV folded light curve.  We
note that the HR is similar to that obtained by \cite{Iaria2001_1822}
and \cite{parmar2000} using BeppoSAX data of X1822-371. The largest
effective area of the Epic-pn allows us to clearly observe a small
variation of the HR, from 1.5 to 1.9, 
which immediately preceeds the eclipse at the
orbital phase 0.949. This feature was already present in the HR shown
by \cite{Iaria2001_1822}, \cite{parmar2000}, and
\cite{Hellier_mason1989}.  Furthermore, we confirm that the depth of
the partial eclipse increases with increasing energy as shown by
\cite{parmar2000}.

\section{Spectral analysis}

\subsection{The averaged spectrum}
\label{avera} 

Chandra data were extracted from regions around the grating arms; to
avoid overlapping between HEG and MEG data, we used a region size of
25 and 33 pixels for the HEG and MEG, respectively, along the
cross-dispersion direction. The background spectra were computed, as
usual, by extracting data above and below the dispersed flux. The
contribution from the background is 0.4\% of the total count rate. We
used the standard CIAO tools to create detector response files
\citep{Davis2001} for the HEG-1 (MEG-1) and HEG+1 (MEG+1) order
(background-subtracted) spectra.  After verifying that the negative
and positive orders were compatible with each other in the whole
energy range, we coadded them using the script {\it
  add$\_$grating$\_$orders} in the CIAO software, obtaining the
first-order MEG spectrum and the first-order HEG spectrum for each of
the two observations.  We summed the first-order HEG (and MEG) spectra
of the two Chandra observations to increase the statistics with the
CIAO script {\it add$\_$grating$\_$spectra} and obtained a total
exposure time of 142 ks. The MEG and HEG energy range is 0.5-6 keV
and 0.8-10 keV, respectively.
Finally, we rebinned the first-order MEG and first-order HEG spectra to have
  at least  25 counts per energy channel. 
\begin{table}[ht]
  \caption{Averaged spectrum. Best-fit values of the continuum parameters.}
\label{Phase_Continuum_averaged}      
\begin{center}
\begin{tabular}{l c c}          
\hline\hline                        
Parameters  &   HETGS & XMM \\

\hline                        

N$_H$  (10$^{22}$ cm$^{-2}$)&  
$0.163$ (fixed)&
$0.163^{+0.008}_{-0.006}$\\
 & & \\

N$_{H_{pc}}$  (10$^{22}$ cm$^{-2}$)&   
$4.84 \pm 0.10$&
$4.97 \pm 0.12  $\\

$f$ &
 $0.654 \pm 0.012$&
$0.607 \pm 0.009$\\
& & \\

kT$_{bb}$ (keV)&
0.061 (fixed) &
$0.061 \pm 0.003$\\

N$_{bb}$ ($\times 10^{-3}$)&
$1.5 \pm 1.0$ &
$1.0 \pm 0.2$\\

 & & \\

kT$_{0}$ (keV)&
0.061 (fixed)&
$0.061 \pm 0.003$\\

kT$_{e}$ (keV)&
$3.07^{+0.18}_{-0.14}$&
$3.05 \pm 0.04$\\

$\tau$          &
$19.1 \pm 0.7 $&
$21.0 \pm 0.4  $\\

N$_{CompTT}$ ($\times 10^{-2}$)&
$12.7 \pm 0.4$&
$7.7 \pm 0.2$\\

 & & \\

$\chi_{red}^2(d.o.f.)$ &
$0.82(6125)$  &
$1.06(2400)$  \\

\hline                                             
\end{tabular}
\end{center}

{\small \sc Note} \footnotesize--- Uncertainties are at the  90\% 
c. l. for each parameter.   
The parameters are defined as in XSPEC. The model is discussed in the text.
The equivalent hydrogen column of the local neutral matter is indicated with 
N$_{H_{pc}}$, the covered fraction of the emitting surface is indicated with 
$f$.  The  $\chi_{red}^2(d.o.f.)$ values are obtained taking into account 
the emission  lines reported in Table \ref{Averaged_line}.
 \end{table}

\begin{figure}
\includegraphics[height=8.5cm]{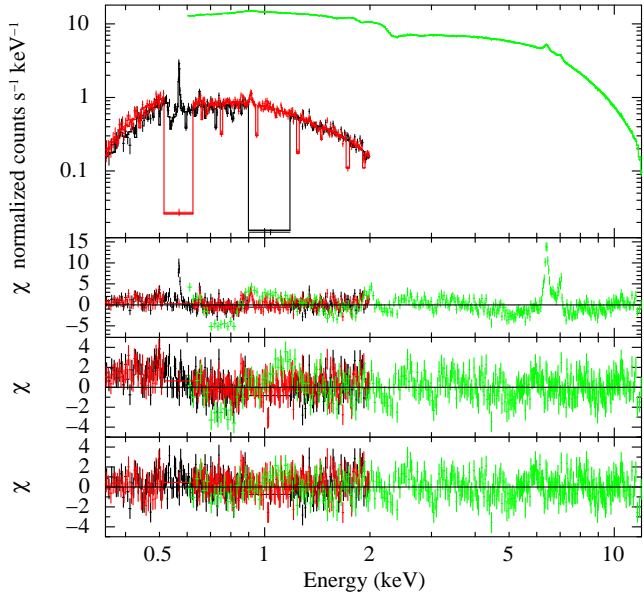}
\caption[]{Data and residuals of RGSs and Epic-pn spectra. Second
  panel from the top: residuals using the model {\tt
    phabs*(f*cabs*phabs*(CompTT)+(1-f)*CompTT)}.  Third panel from the
  top: residuals adding nine Gaussian components to the previous model
  to fit the lines.  Bottom panel: residuals adding a black-body
  component to the model.  }
\label{Fig_res_aver_XMM}
\end{figure}

\begin{figure}
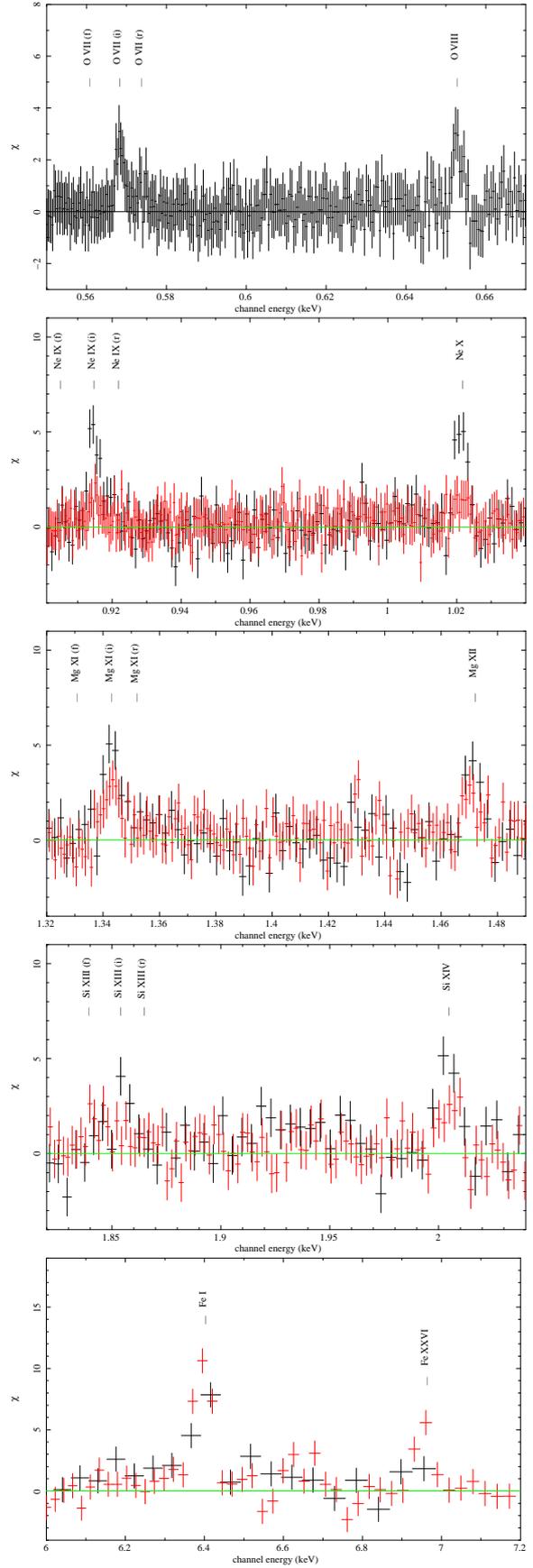

\includegraphics[height=7.5cm,angle=0, angle=-90]{fig6a.ps}\\
\includegraphics[height=7.5cm,angle=0, angle=-90]{fig6b.ps}\\
\includegraphics[height=7.5cm,angle=0, angle=-90]{fig6c.ps}\\
\includegraphics[height=7.5cm,angle=0, angle=-90]{fig6d.ps}\\
\includegraphics[height=7.5cm,angle=0, angle=-90]{fig6e.ps}
\caption[]{Residuals of first-order MEG (black points) and HEG (red points) 
spectra using the continuum 
shown in the text. The emission lines  are indicated in the panels.}
\label{Fig_res_aver_Ch}
\end{figure}  

\begin{table*}[ht]
  \caption{Averaged spectrum.  Emission lines of X1822-371.}
\label{Averaged_line}      
\begin{center}
\begin{tabular}{l c c c c c c c }          
\hline\hline 
 &               
Identification  &
Expected  &
E &
$\sigma$& 
I  &
eq. width  & 
significance \\
 & &(keV)&(keV) & (eV) &  (10$^{-4}$ ph cm$^{-2}$ s$^{-1}$)&
(eV)&  ($\sigma$)\\

\hline                        
\multirow{10}{*}{HETGS}& \ion{O}{vii} (i) &
0.5687&
 $ 0.5684 \pm 0.0003$& 
$1.1 \pm 0.2$&
$34 \pm 6$&
$19 \pm 6$ &
6  \\

&\ion{O}{viii} & 
0.6536   &
 $ 0.6529^{+0.002}_{-0.004} $& 
$1.1 \pm 0.3$&
$9 \pm 2$&
$6 \pm 2$ &
5  \\

&\ion{Ne}{ix} (i)& 
0.9149 &
$ 0.9147 \pm 0.0003$& 
$1.7 \pm 0.3$&
$4.3 \pm 0.5$&
$5.2 \pm 1.0$ &
9  \\

&\ion{Ne}{x} &
1.0218 &
$ 1.0212 \pm 0.0003$& 
$1.74^{+0.14}_{-0.30}$&
$2.4 \pm 0.3$&
$3.4 \pm 0.8$ &
8 \\

&\ion{Mg}{xi} (i)&
1.3433&
$ 1.3430 \pm 0.0004$ &
$2.8 \pm 0.5$&
$1.40 \pm 0.14$&
$2.8 \pm 0.6$ &
10 \\

&\ion{Mg}{xii} &
1.4723&
$ 1.4706 \pm 0.0005$ &
$2.5 \pm 0.4 $&
$0.82 \pm 0.10$   &
$1.9 \pm 0.6$ &
8  \\

&\ion{Si}{xiii} (i) & 
1.8542&
$ 1.8551 \pm 0.0010$ &
3 (fixed)  &
$0.46 \pm 0.08$  &
$1.4 \pm 0.6$ &
6   \\

&\ion{Si}{xiv} & 
2.005&
$ 2.0043 \pm 0.0007$ &
$3.7 \pm 0.6$ &
$0.74 \pm 0.09$  &
$2.5 \pm 0.7$ &
 8 \\

&\ion{Fe}{i} &
6.404 &
$ 6.393 \pm 0.002$ &
$18 \pm 3$ &
$3.2 \pm 0.2$  &
$40 \pm 6$ &
16 \\

&\ion{Fe}{xxv}(i) &
6.675&
6.675 (fixed) &
$13$ (fixed) &
$0.6 \pm 0.2$  &
$8 \pm 4$ &
3 \\

&\ion{Fe}{xxvi}  &
6.966&
$ 6.952 \pm 0.004$ &
$13 \pm 6$ &
$1.5 \pm 0.2$  &
$22 \pm 8$ &
8 \\
\\

\multirow{5}{*}{RGS}&\ion{O}{vii} (i) & 
0.5687&
$ 0.56869^{+0.00011}_{-0.00016}$ &
$1.4 \pm 0.2$ &
$23.2 \pm 1.3$  &
$22 \pm 2 $  &
18 \\

&\ion{O}{viii} & 
0.6536   &
 $ 0.651794^{+0.000336}_{-0.000004} $& 
$<0.7$&
$3.6 \pm 0.4 $  &
$4.6 \pm 1.0$  &
9  \\

&\ion{Ne}{ix} (i) & 
0.9149   &
 $ 0.9139 \pm 0.0011 $& 
$4.4 \pm 1.3$&
$3.8 \pm 0.3$  &
$8.0 \pm 1.3 $  &
13  \\

&\ion{Ne}{x} & 
1.0218   &
1.0218 (fixed)  & 
2 (fixed)&
$1.3 \pm 0.2$  &
$3.2 \pm 1.1$  &
7  \\

&\ion{Mg}{xi} (i)&
1.3433&
1.3433 (fixed)&
3 (fixed)&
$0.6 \pm 0.2$&
$2.0 \pm 1.1$ &
3 \\
\\

\multirow{4}{*}{Epic-pn}

&\ion{Si}{xiv} & 
2.005&
$ 1.984^{+0.010}_{-0.006} $ &
4 (fixed)&
$1.49 \pm 0.15$ &
$ 8 \pm 2$ &
10 \\

&\ion{Fe}{i} &
6.404 &
$ 6.402 \pm 0.004$ &
$75 \pm 8 $ &
$3.24 \pm 0.13$ &
$ 56 \pm 5$  &
25 \\

&\ion{Fe}{xxv} (i)&
6.675 &
$6.703 \pm 0.013$  &
20 (fixed) &
$0.70 \pm 0.08$  &
$12 \pm 3$  &
9 \\

&\ion{Fe}{xxvi}  & 
6.966&
$ 6.991 \pm 0.009$ &
$56^{+8}_{-16} $ &
$1.38^{+0.11}_{-0.06}$  &
$26 \pm 5$  &
23  \\

\hline                                             
\end{tabular}
\end{center}

{\small \sc Note} \footnotesize--- Best-fit values of the HETGS and XMM
emission lines  in X1822-371.  
Uncertainties are  at the  68\% c. l. 
for each parameter. In the first column we report the 
atomic transition, from the second to the fifth column 
the expected rest-frame energy, the centroid, 
the width, and the intensity of each line.  In 
the sixth and seventh column we report the equivalent width 
and the significances,  in units of $\sigma$,  of 
each line.  
\end{table*}

We extracted the RGS1, RGS2, and Epic-pn spectra from the whole
observation with an exposure time of 43, 42, 38 ks, respectively. We
adopted a 0.35-2 keV energy range for RGS spectra and 0.6-12 keV for
the Epic-pn spectrum.  The Epic-pn spectrum was rebinned so that the
energy resolution was not oversampled by more than a factor 4 and to
have at least 25 counts per energy channel.  The RGS1 and RGS2 spectra
were rebinned to have at least 25 counts per energy channel.

The HETGS and XMM spectra were fitted separately using XSPEC version
12.6.0. We fitted the continuum emission adopting the model proposed
by \cite{Iaria2001_1822} for the broad band BeppoSAX spectrum of
X1822-371: i.e.  a Comptonised component ({\tt CompTT} in XSPEC)
partially absorbed by neutral matter and absorbed by interstellar
matter ({\tt phabs} in XSPEC, using the abundances of \cite{aspl}).
We also took into account the effect of Thomson scattering in the
local cold absorber using the {\tt cabs} component in XSPEC.  The
adopted model consists of the component {\tt
  phabs*(f*cabs*phabs*(CompTT)+(1-f)*CompTT)}, where f is the neutral
matter covering fraction; the first {\tt phabs} component was used to
fit the photoelectric absorption by interstellar neutral matter, the
second {\tt phabs} component takes into account the photoelectric
absorption by local neutral matter. We imposed that the value of the
equivalent hydrogen column density of the {\tt cabs} component is the
same as that of the neutral local matter, assuming that the local
neutral matter is responsible for both the photoelectric absorption
and Thomson scattering.

Fitting the RGS and Epic-pn spectra, we obtain a $\chi_{\rm
  red}^2({\rm {d.o.f.}})$ of 1.86(2422).  The residuals reported
  in Fig.  \ref{Fig_res_aver_XMM} (the second panel from the top) show
   several   local features and a mismatch between
  the RGS and Epic-pn data in the energy range between 0.6 and 0.8
  keV.  We improved the model by adding nine Gaussian components to fit
the localised features.  The centroids of the Gaussian profiles are
at 0.569, 0.652, 0.914, 1.02, 1.34, 1.98, 6.40, 6.70, and 6.99 keV,
associated with the \ion{O}{vii} intercombination line, \ion{O}{viii},
\ion{Ne}{ix} intercombination line, \ion{Ne}{x}, \ion{Mg}{xi}
intercombination line, \ion{Si}{xiv}, \ion{Fe}{i}, \ion{Fe}{xxv}
intercombination line, and \ion{Fe}{xxvi}, respectively.  The addition
of the Gaussians improved the fit, giving a $\Delta \chi^2$
of 1691 and a $\chi_{\rm red}^2({\rm {d.o.f.}})$ of 1.17(2401); 
  the corresponding residuals are shown in Fig. \ref{Fig_res_aver_XMM}
  (the third panel from the top). To improve the fit at low energies,
  where  RGS and Epic-pn data are mismatched, we added a
  black-body component ({\tt bbody} in XSPEC) to our model, partially
  absorbed like the Comptonised component. Since the best-fit gives a
  value of 0.06 keV for the black-body temperature and for the
  seed-photon temperature of the CompTT component, we constrained the
  two parameters to assume the same value and fitted the data
  again. The addition of the black-body component improved the fit,
  giving a $\Delta \chi^2$ of 250 and a $\chi_{\rm
    red}^2({\rm{d.o.f.}})$ of 1.06(2400). We report the corresponding
  residuals in Fig. \ref{Fig_res_aver_XMM} (bottom panel). The
  best-fitting values of the parameters of the continuum are reported
  with an associated error at 90\% confidence level (c.l.) in Table
  \ref{Phase_Continuum_averaged} and the best-fitting values of the
  line parameters are reported in Table \ref{Averaged_line}
  (associated error at 68\% c. l.).

We fitted the HETGS data with the same best-fit continuum as adopted
above, obtaining a value of $\chi_{\rm red}^2({\rm
  {d.o.f.}})=0.95(6155)$; we observed several emission features in the
residuals as shown in Fig.  \ref{Fig_res_aver_Ch}. To fit the
residuals, we added eleven Gaussian components and identified the
emission lines associated with the following transitions: \ion{O}{vii}
intercombination line, \ion{O}{viii} Ly$\alpha$, \ion{Ne}{ix}
intercombination line, \ion{Ne}{x} Ly$\alpha$, \ion{Mg}{xi}
intercombination line, \ion{Mg}{xii} Ly$\alpha$, \ion{Si}{xiii}
intercombination line, \ion{Si}{xiv} Ly$\alpha$, \ion{Fe}{i}
K$\alpha$, \ion{Fe}{xxv} intercombination line, and, finally,
\ion{Fe}{xxvi} Ly$\alpha$.   The addition of the Gaussian
components improved the fit with a $\Delta \chi^2$ of 807 and a value
of $\chi_{\rm red}^2({\rm {d.o.f.}})$ of 0.82(6125).  The best-fit
  values of the continuum parameters and the best-fit values of the
  line parameters are reported in Tables \ref{Phase_Continuum_averaged}
  and \ref{Averaged_line}, respectively.

 The black-body component added to fit the XMM spectrum below 0.8
  keV is not present in the model shown by \cite{Iaria2001_1822}. This
  component was also added to fit the Chandra averaged spectrum, fixing
  its temperature to the best-fit value obtained by the XMM averaged
  spectrum.  We obtained a consistent value for its
  normalisation.

 Finally, we  checked the presence in the RGS spectrum of the
  \ion{Mg}{xii} and \ion{Si}{xiii} lines, which are significantly
  detected in the HETGS spectrum.  We added two Gaussian components in
  the XMM averaged spectrum,  keeping  the centroids and the widths fixed
  to the values obtained from Chandra. We obtained an upper limit to the
  flux of the \ion{Mg}{xii} and \ion{Si}{xiii} line of $<0.12 \times
  10^{-4}$ and $<0.5 \times 10^{-4}$ ph cm$^{-2}$ s$^{-1}$,
  respectively.  These fluxes suggest that the \ion{Mg}{xii} line
  might be absent during the XMM observation while the \ion{Si}{xiii}
  line is compatible with that observed in the Chandra averaged
  spectrum.

\subsection{Helium-like series and the \ion{Ne}{ix} RRC feature}
\label{trip}
The intercombination lines associated with \ion{O}{vii} (RGS1
spectrum), \ion{Ne}{ix} (MEG spectrum), and \ion{Mg}{xi} (MEG spectrum)
have a significance larger than 7 $\sigma$ but we do not have evidence
for the corresponding forbidden and resonance lines.  To determine the
intensity of the \ion{O}{vii} lines, we analysed the energy range
between 0.555 and 0.58 keV, fixing the best-fitting values of the
continuum emission. We studied the \ion{O}{vii} triplet using RGS1 and
MEG data.  In the MEG and RGS1 spectra we fixed the energy of the
  forbidden line at the rest-frame value of 0.5610 keV, while the
  energies of the intercombination and resonance lines were let free to
  vary.  The line widths were constrained to assume the same value.
\begin{figure}
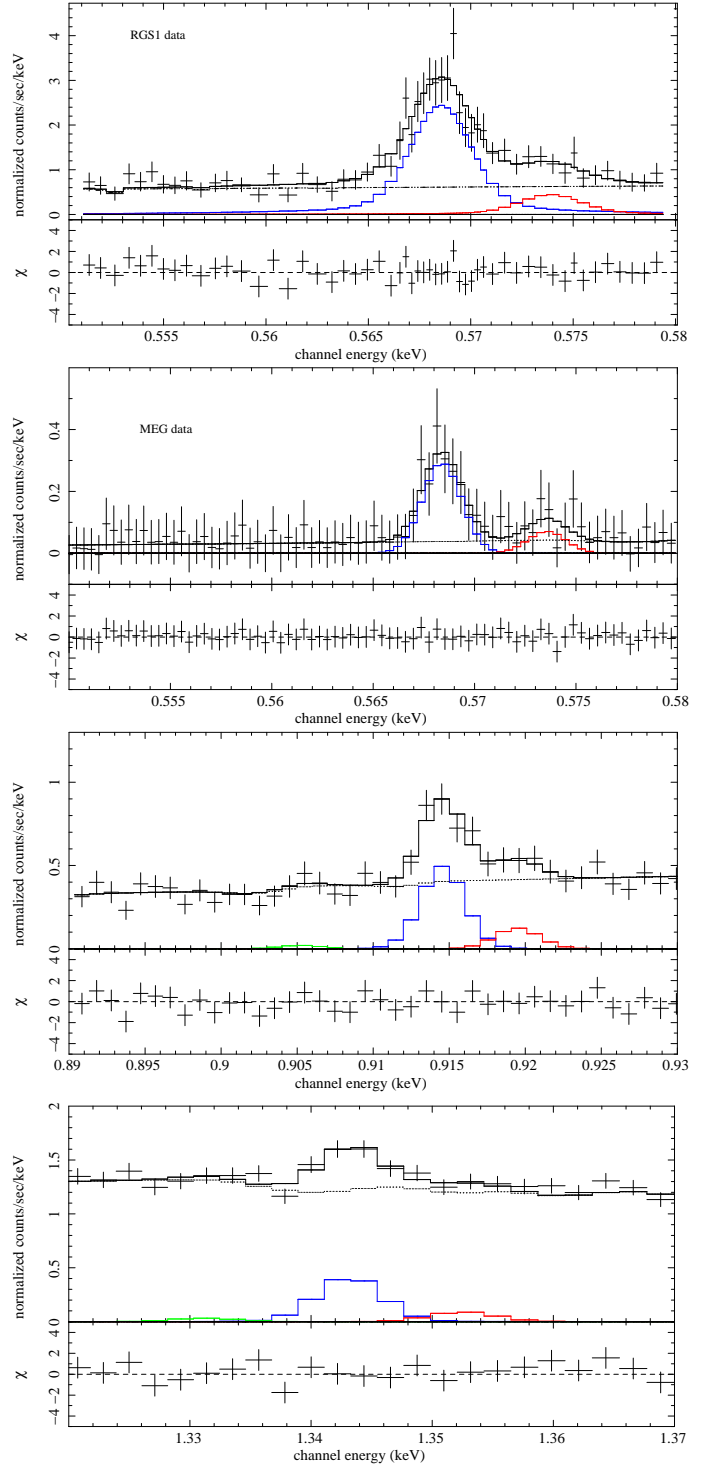

\includegraphics[height=4.8cm, angle=0]{fig7a.ps}\\
\includegraphics[height=4.8cm, angle=0]{fig7b.ps}\\
\includegraphics[height=4.8cm, angle=0]{fig7c.ps}\\
\includegraphics[height=4.9cm, angle=0]{fig7d.ps}
\caption[]{Forbidden (green), intercombination 
(blue), and resonance (red) lines of the  \ion{O}{vii}, 
 \ion{Ne}{ix} and  \ion{Mg}{xi} triplets. From top to bottom: 
 RGS1 data and residuals  of the \ion{O}{vii} triplet; first-order MEG 
data and residuals of the \ion{O}{vii} triplet;  first-order  MEG
data and residuals of the \ion{Ne}{ix} triplet;   first-order MEG
data and residuals of the \ion{Mg}{xi} triplet. }
\label{Fig_res_oxy}
\end{figure}
Using the best-fitting intensities reported in Table
\ref{Averaged_triplet}, we derived the parameter $G=(f+i)/r$ and
$R=f/i$, where $f$, $i$, and $r$ are the intensities of the forbidden,
intercombination, and resonance lines, respectively.  We find that
$G=4.3^{+2.9}_{-1.5}$ and $G=7 \pm 4$ for MEG and RGS1 data and $R$ is
less than 0.07 and 0.02 for MEG and RGS1, respectively (the associated
errors are at 68\% c. l.).  We used the MEG data between 0.89 and
0.93 keV to estimate $G$ and $R$ of the \ion{Ne}{ix} triplet.  We
fixed the forbidden line energy at the rest-frame value of 0.9051 keV
and imposed the widths of the three lines to be the same, finding
$G=4.4^{+3.9}_{-1.3}$ and $R=0.04^{+0.08}_{-0.04}$.  Finally, we
used the MEG data between 1.32 and 1.37 keV to determine $G$ and
$R$ of the \ion{Mg}{xi} triplet,  we fixed the energies of the
  forbidden and resonance line at the rest-frame values of 1.3312
  and 1.3523 keV, while the line widths were constrained to assume
the same value. We obtained $G=3.7^{+3.7}_{-1.5}$ and
$R=0.11^{+0.14}_{-0.09}$.

\begin{table}[h]
  \caption{Best-fit values of the \ion{O}{vii}, \ion{Ne}{ix}, and 
\ion{Mg}{xi} triplets.}
\label{Averaged_triplet}      
\begin{center}
\begin{tabular}{c c c c c}          
\hline\hline 

& &E &
$\sigma$& 
I  \\

& &(keV)& (eV) &  (10$^{-4}$ ph cm$^{-2}$ s$^{-1}$)\\

\hline
\multicolumn{5}{c}{\ion{O}{vii}}\\

\multirow{3}{*}{MEG}&f & $0.5610$ (fixed)&
$1.0 \pm 0.2$ &
$<2$\\

&i & $0.5684 \pm 0.0002$& 
$1.0 \pm 0.2$ &
$26 \pm 3$ \\

&r & $0.5738 \pm 0.0010$&
$1.0 \pm 0.2$ &
$6 \pm 2$\\
 & & & & \\
\multirow{3}{*}{RGS1}&f & $0.5610$ (fixed) &
$1.0 \pm 0.2$  &
$<0.2$\\
&i & $0.56850^{+0.00009}_{-0.00020}$ & 
$1.0 \pm 0.2$ &
$12.6 \pm 1.3$ \\
&r & $0.5725^{+0.0019}_{-0.0011}  $ &
$1.0 \pm 0.2$ &
$1.9^{+1.0}_{-0.8}$\\
\multicolumn{5}{c}{\ion{Ne}{ix}}\\
\multirow{3}{*}{MEG}&f &$0.9051$ (fixed) &
$1.3 \pm 0.2$ &
$0.12^{+0.25}_{-0.12}$\\
&i & $0.9145 \pm 0.0002$& 
$1.3 \pm 0.2$ &
$3.4 \pm 0.4$ \\
&r & $0.9194 \pm 0.0008$&
$1.3 \pm 0.2$ &
$0.8 \pm 0.3$\\
\multicolumn{5}{c}{\ion{Mg}{xi}}\\
\multirow{3}{*}{MEG}&f &$1.3312$ (fixed)  &
$2.5 \pm 0.4$ &
$0.12 ^{+0.12}_{-0.10}$\\
&i & $1.3429 \pm 0.0004$&
$2.5 \pm 0.4$ &
$1.10 \pm 0.14$\\
&r &  1.3523 (fixed) &
$2.5 \pm 0.4$ &
$0.33 ^{+0.11}_{-0.13}$\\
\hline                                             
\end{tabular}
\end{center}

{\small \sc Note} \footnotesize---  
Centroid,  width, and  intensity of the
 He-like transition lines. The errors associated with each 
parameter are at 68\%  c. l.. 
\end{table}

\cite{Cottam2001} found a radiative recombination continuum (RRC)
feature associated with the \ion{Ne}{ix} transition in a previous
Chandra/HETGS observation of X1822-371. They found an energy threshold
of $13 \pm 7$ eV, corresponding to a plasma temperature of $(1.5 \pm
0.8) \times 10^5$ K.  We  searched for this feature in our data by
adding an inverse edge component ({\tt redge} in XSPEC)  to assess
the presence of the RRC feature. We  fixed the energy threshold at
1.196 keV (that is the energy of the \ion{Ne}{ix} transition) and
found only marginal evidence for the RRC feature ($\Delta \chi^2 =
2.8$), with a temperature $<$14 eV corresponding to  a temperature of
$<$1.6$ \times 10^5$ K.

\subsection{Phase-dependent spectral analysis}
\label{analysis}

We selected the following five phase-intervals 0.04-0.25, 0.25-0.50,
0.50-0.75, 0.75-0.95, and 0.95-1.04, respectively. This selection
allows us to constrain both continuum emission and discrete features for
each of phase-interval. The  intervals correspond to the
count rate rise after the eclipse, the maximum, the count rate
decrease, the minimum and, finally, the eclipse, respectively, as
shown in Fig. \ref{Fig2}.

\begin{table*}[ht]
  \caption{Exposure times of the selected phase-intervals.}
\label{Tab1}      
\begin{center}                                      
\begin{tabular}{l c c c c c }          
\hline\hline 
Phase interval&
Obs. 9076&
Obs. 9858&
RGS1 &
RGS2&
Epic-pn
\\
\hline         
0.04-0.25& 
15.0 ks&
17.2 ks&
8.1 ks &
7.9 ks &
7.3 ks 
\\
   
0.25-0.50&
14.7 ks&
20.8 ks&
9.8 ks&
9.5 ks&
13 ks\\
 
0.50-0.75&
14.7 ks&
19.6 ks&
12.8 ks &
12.4 ks &
8.5 ks \\

0.75-0.95&
11.7 ks&
15.7 ks&
7.9 ks&
7.6 ks&
7.2 ks\\
 
0.95-1.04&
6.1 ks&
7.0 ks& 
3.6 ks&
3.6 ks&
1.8 ks\\

\hline\hline 

\end{tabular}
\end{center}                                      

{\small \sc Note} \footnotesize---  We show  the 
selected 
phase-intervals in the first column,  the 
corresponding exposure times of the two HETGS spectra 
in the second and third columns, the exposure 
times of the RGS1, RGS2, and Epic-pn spectra in the fourth, 
fifth, and sixth columns, respectively.        
 \end{table*}

\begin{table*}[ht]
  \caption{Best-fit values of the continuum parameters as a function of 
 the orbital  phase of   X1822-371}
\label{Phase_Continuum}      
\begin{center}
\begin{tabular}{l c c c c c c }          
\hline\hline                        
Spectra& Parameters  &
 \multicolumn{5}{c}{Phase intervals}  \\
& & 0.04-0.25&  0.25-0.50&  0.50-0.75&  0.75-0.95 
& 0.95-1.04\\
& & 1&  2&  3&  4 
& 5\\

\hline                        
&&&&&&\\
\multirow{10}{*}{HETGS} &N$_H$  (10$^{22}$ cm$^{-2}$)&  
0.163 (fixed)&0.163 (fixed)&0.163 (fixed)&0.163 (fixed)&
0.163 (fixed)\\

 & & & & & & \\
 
 &N$_{H_{pc}}$  (10$^{22}$ cm$^{-2}$)&   
$4.9 \pm 0.2 $&
$4.8 \pm 0.2 $&
$4.7 \pm 0.3 $&
$4.8 \pm 0.3 $&
$4.6^{+0.8}_{-0.6} $\\
&$f$ &
 $0.678^{+0.014}_{-0.027}$&
$0.63 \pm 0.03$&
$0.65 \pm 0.04$&
$0.66 \pm 0.04$&
$0.65^{+0.06}_{-0.09}$\\
 & & & & & & \\

&kT$_{0}$ (keV)&
0.060 (fixed)&
0.059 (fixed)&
0.054 (fixed)&
0.054 (fixed)&
0.059 (fixed)\\

&kT$_{e}$ (keV)&
$3.31 ^{+0.31}_{-0.07}$&
$3.2 ^{+0.5}_{-0.3}$&
$2.8 \pm 0.3 $&
$3.3^{+1.0}_{-0.4} $&
$3.4^{+71.0}_{-0.8} $\\

&$\tau$          &
$17.80 \pm 0.14 $&
$20 \pm 2  $&
$20 \pm 2   $&
$18 \pm 2  $&
$18^{+4}_{-12}$\\

&N$_{Compst}$ ($10^{-2}$)&
$13.6^{+0.4}_{-0.7}$&
$17.8 \pm 0.8 $&
$13.2 \pm 0.9$&
$10.7^{+1.0}_{-1.3} $&
$8.6^{+13.5}_{-8.5}$\\
 & & & & & & \\
&Unabs. Flux&

$9.61^{+0.06}_{-0.14}$&
$10.85^{+0.06}_{-0.14}$&
$8.37^{+0.05}_{-0.15}$&
$7.63^{+0.07}_{-0.11}$&
$6.32^{+0.09}_{-0.18}$\\
& & & & & & \\

&$\chi_{red}^2(d.o.f.)$ &
$0.70(3687)$  &
$0.73(4035)$  &
$0.67(3611)$  &
$0.63(3029)$  &
$0.49(1467)$  \\

\hline& & & & & & \\

\multirow{10}{*}{XMM} &N$_H$  (10$^{22}$ cm$^{-2}$)&  
0.163 (fixed) &
0.163 (fixed) &
0.163 (fixed) &
0.163 (fixed) &
0.163 (fixed) \\

& & & & & & \\

&
N$_{H_{pc}}$  (10$^{22}$ cm$^{-2}$)&  
$5.1 \pm 0.2  $&
$5.5 \pm 0.2  $&
$5.1 \pm 0.2  $&
$4.4 \pm 0.2 $&
$5.1 \pm 0.4 $\\

&$f$          &
$0.604^{+0.011}_{-0.022}  $&
$0.612 \pm 0.014 $&
$0.643^{+0.013}_{-0.017}  $&
$0.58^{+0.02}_{-0.03}   $&
$0.63^{+0.04}_{-0.05} $\\
& & & & & & \\

&kT$_{bb}$ (keV)         &
$0.060^{+0.004}_{-0.007}  $&
$0.059 \pm 0.006 $&
$0.054 \pm 0.005$&
$0.054^{+0.007}_{-0.005}$&
$0.059 \pm 0.015 $\\

&N$_{bb}$    ($\times 10^{-3}$)     &
$1.0^{+0.8}_{-0.3}$    &
$1.2^{+0.8}_{-0.4}$    &
$1.4 \pm 0.7$    &
$1.6^{+1.1}_{-0.7}$    &
$0.7^{+1.7}_{-0.5}$    \\

& & & & & & \\

&kT$_{0}$ (keV)         &
$0.060^{+0.004}_{-0.007}  $&
$0.059 \pm 0.006 $&
$0.054 \pm 0.005$&
$0.054^{+0.007}_{-0.005}$&
$0.059 \pm 0.015 $\\

&kT$_{e}$ (keV)         &
$3.06 \pm 0.09 $&
$3.11 \pm 0.07 $&
$3.06 \pm 0.07$&
$2.98 \pm 0.08$&
$3.0 \pm 0.2 $\\
 
&$\tau$          &
$20.3 \pm 0.7$&
$20.6 \pm 0.5$&
$20.0 \pm 0.6 $&
$22.5 \pm 0.8 $&
$20.0 ^{+1.7}_{-1.5}$\\
 
&N$_{Compst}$ ($\times 10^{-2}$)&
$7.7^{+0.5}_{-0.3} $&
$8.8 \pm 0.5$&
$8.99 \pm 0.12 $&
$6.5 \pm 0.3 $&
$5.4 \pm 0.7 $\\
& & & & & & \\

&Unabs. Flux&
$5.99 \pm 0.03$&
$7.22 \pm 0.03$&
$6.60 \pm 0.03$&
$5.52 \pm 0.03$&
$4.05^{+0.05}_{-0.03}$\\
& & & & & & \\

&$\chi_{red}^2(d.o.f.)$ &
$1.00(839)$ &
$1.06(1015)$&
$0.97(1115)$ &
$1.01(728)$ &
$0.98(487)$ \\

\hline                                             
\end{tabular}
\end{center}

{\small \sc Note} \footnotesize---Uncertainties are at the  90\% 
c. l. for each parameter. The 0.4-12 keV unabsorbed flux 
is given in units of $10^{-10}$ erg s$^{-1}$ cm$^{-2}$. The 
 $\chi_{red}^2(d.o.f.)$ 
and $\chi^2(d.o.f.)$ values are obtained taking into account 
the emission  lines shown in Tab. \ref{Line1}.  
\end{table*}

For each phase-interval we produced a spectrum using the same
procedure as adopted for the averaged spectrum.  We  label 
 the spectra extracted from the phase-intervals 0.04-0.25,
0.25-0.50, 0.50-0.75, 0.75-0.95, and 0.95-1.04  with  1, 2,
3, 4, and 5, respectively. The
exposure times of the spectra corresponding to the five
phase-intervals are reported in Table \ref{Tab1}.  The MEG, HEG,
  RGS, and EPIC-pn selected energy ranges are the same as adopted to fit
  the averaged spectrum.  The model to fit the XMM spectra is
  the same as used to fit the averaged spectrum, while we fitted the
  five Chandra spectra excluding the black-body component because of
  the low statistics below 0.6 keV in the MEG.  Furthermore,
  we fixed the $N_H$ value to $0.163 \times 10^{22}$ cm$^{-2}$ in the
  Chandra and XMM spectra and, in the Chandra spectra, the seed photon
  temperature of the Comptonised component was fixed to the values
  obtained from the best-fit of the corresponding  XMM spectrum.

\begin{figure}
\includegraphics[height=7.5cm, angle=0]{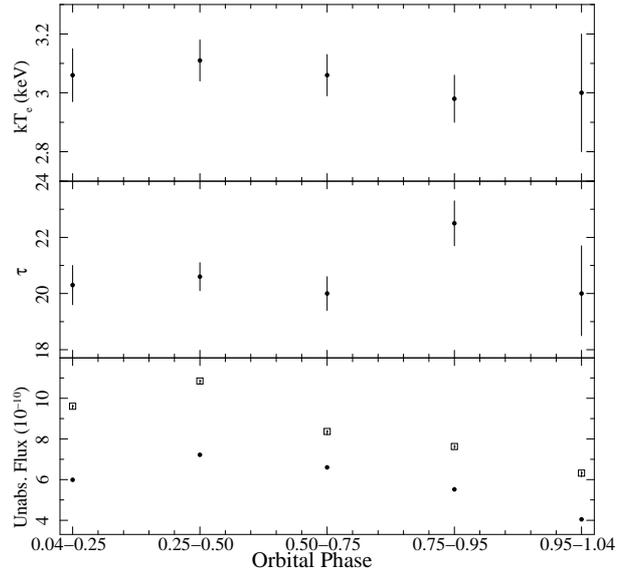}
\caption[]{Parameters of the Comptonised component obtained
  fitting the XMM data: electron temperature (top), optical depth
  (middle), and the unabsorbed flux in units of $10^{-10}$ erg s$^{-1}$
  cm$^{-2}$ (bottom) in the 0.4-12 keV energy band, the open squares
  and the dots are the values obtained from the Chandra and XMM
  spectra, respectively.   }
\label{Fig_cont}
\end{figure}

  We obtained acceptable fits, the XMM spectra having 
  $\chi_{red}^2(d.o.f.)$ of $1.00(815)$, $1.04(982)$, $0.98(1080)$,
  $1.00(707)$, and $0.97(475)$ for spectra 1, 2, 3, 4, and 5,
  respectively.  The best-fit parameters are shown in Tables
  \ref{Phase_Continuum} and \ref{Line1}. The electron temperature
  $kT_e$ does not change with the orbital phase (see top panel of
    Fig. \ref{Fig_cont}), while the optical depth $\tau$ of the
    Comptonised component is about 20 at all  orbital phases
    except between 0.75-0.95, when it increases up to 23 (see middle
    panel of Fig. \ref{Fig_cont}).  We show the unabsorbed flux in
  the 0.4-12 keV energy range associated with the Comptonised
  component that was obtained using the {\tt cflux} component of XSPEC in
  the bottom panel of Fig.  \ref{Fig_cont}, where the open squares and
  the dots indicate the Chandra and XMM unabsorbed fluxes,
  respectively.  The behaviour of the unabsorbed fluxes obtained from
  the XMM and Chandra phase-resolved spectra is similar along the
  orbit. The Chandra unabsorbed fluxes are a factor 1.4 higher than
  the those of XMM in all selected phase-intervals.

\begin{figure}[ht]
\includegraphics[height=7.5cm, angle=0]{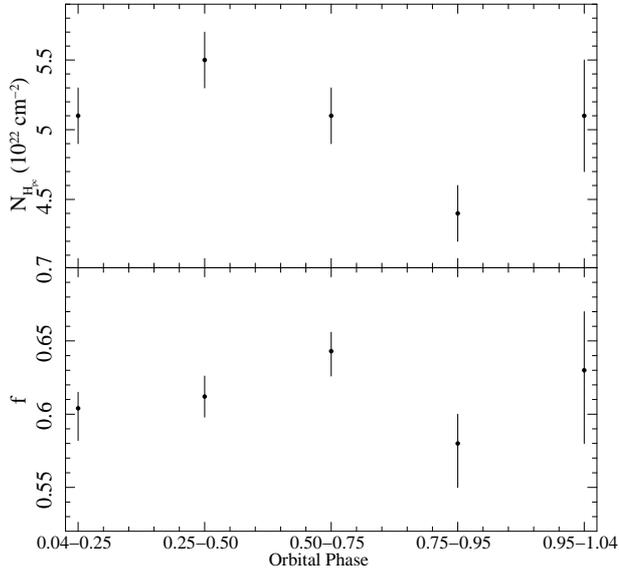}
\caption[]{Best-fit values of the equivalent hydrogen
  column density associated with the local neutral matter (upper
  panel) and the fraction $f$ of the emitting region covered by local
  neutral matter (lower panel).  }
\label{pcfainit}
\end{figure}

  The equivalent hydrogen column associated with the local neutral
  matter, $N_{H_{pc}}$, reaches the maximum at phases between
  0.25-0.50 when the unabsorbed flux is maximum. It assumes the
  minimum value at phases between 0.75-0.95 (see Fig \ref{pcfainit},
  upper panel).  A similar behaviour of $N_{H_{pc}}$ and $f$ was
  already found by \cite{Heinz_nowak_2001} out of eclipse from analysing
  ASCA and RXTE data simultaneously (see Table 2 in their
  work). However, we found that the values of $N_{H_{pc}}$ and $f$
  increase again during the eclipse, while \cite{Heinz_nowak_2001}
  found a further drop.

\begin{table*}[ht]
  \caption{Emission lines of X1822-371}
\tiny
\label{Line1}      
\begin{center}
\begin{tabular}{ c c c c c c c  c c c c}          
\hline\hline 
&Id.  &
E &
$\sigma$& 
I(10$^{-4}$)  &
significance  & 
E &
$\sigma$& 
I(10$^{-4}$) &
significance \\

&  &(keV) & (eV) &  (ph cm$^{-2}$ s$^{-1}$)& ($\sigma$) 
  &(keV) & (eV) &  (ph cm$^{-2}$ s$^{-1}$)& ($\sigma$)\\

\hline 
 & & & & & & & \\
    
        \multicolumn{2}{c}{ } &
\multicolumn{4}{c}{0.04-0.25 Phase-Interval} &
\multicolumn{4}{c}{0.25-0.50 Phase-Interval} \\

\multirow{9}{*}{HETGS}&
\ion{O}{vii} (i)& 
$ 0.5690^{+0.0007}_{-0.0015}$& 
1.1 (fixed)&
$57 \pm 23 $ &
2.5&
$ 0.5688^{+0.0029}_{-0.0011}$& 
1.1 (fixed)&
$33^{+27}_{-20}$ &
1.7 \\

&\ion{O}{viii} & 

$ 0.6525 \pm 0.0008$& 
$1.3 \pm 0.5 $&
$15 \pm 6$ &
2.5&
$ 0.6538 \pm 0.0012$& 
1.3 (fixed)&
$10 \pm 5$&
2  \\

&\ion{Ne}{ix} (i)& 

$ 0.9143^{+0.0002}_{-0.0004}$& 
$0.9 \pm 0.3$&
$6.0 \pm 1.2$ &
5&
$ 0.9159 \pm 0.0005$& 
$1.8 \pm 0.5$&
$6.1 \pm 1.3$&
4.7 \\

&\ion{Ne}{x} &

$ 1.0207 \pm 0.0004$& 
$1.2\pm 0.4 $&
$3.4 \pm 0.8$ &
4.3&
$ 1.0216 \pm 0.0004$& 
$1.7 \pm 0.4$&
$3.4 \pm 0.7 $&
4.9 \\

&\ion{Mg}{xi} (i)&

$ 1.3423 \pm 0.0004$ &
$1.2 \pm 0.5$&
$1.5 \pm 0.3$ &
5&
$ 1.3440 \pm 0.0006$ &
$3.2^{+1.0}_{-0.7}$&
$2.4 \pm 0.4$ &
6\\

&\ion{Mg}{xii} &

$ 1.4705 \pm 0.0004$ &
$1.3 \pm 0.5$&
$1.2 \pm 0.2$  &
6&
$ 1.4725 \pm 0.0009$ &
$2.2^{+1.3}_{-0.9} $&
$0.9 \pm 0.3$ &
3  \\

&\ion{Si}{xiv} & 

$ 2.0038 \pm 0.0008$ &
$2.4^{+1.5}_{-0.8}$ &
$1.1 \pm 0.2$  &
5.5&
$ 2.0071 \pm 0.0011 $ &
$2.8 \pm 1.0$ &
$0.8 \pm 0.2$ &
4 \\

&\ion{Fe}{i} &

$ 6.396 \pm 0.003$ &
$14 \pm 3 $ &
$4.1 \pm 0.5$  &
8.2&
$ 6.396 \pm 0.004$ &
$19 \pm 5 $ &
$4.4 \pm 0.6$ &
7.3\\

&\ion{Fe}{xxvi}  &

$ 6.960 \pm 0.007$ &
$<22$ &
$2.1 \pm 0.5$  &    
4.2&
$ 6.946^{+0.008}_{-0.002} $ &
 $<9$&
$1.8 \pm 0.5$  &
3.6 \\

& & & & & & & \\

\multirow{6}{*}{XMM}&
\ion{O}{vii} (i) & 
$ 0.5684 \pm 0.0003$ &
$1.6 \pm 0.5$ &
$25 \pm 3 $   &
8.3&
$ 0.5688^{+0.0002}_{-0.0003} $ &
$1.1 \pm 0.3$ &
$28 \pm 3   $ &
9.3 \\

&\ion{O}{viii} & 
$ 0.651 \pm 0.002$& 
$<3$&
$3.0 \pm 1.0$ &
3&
$ 0.6536 \pm 0.0009$&
3 (fixed)&
$5.7 \pm 1.2$&
4.8\\

&\ion{Ne}{ix} (i)& 
$ 0.913 \pm 0.002$& 
 $4 \pm 3$&
$4.3 \pm 0.7$ &
6.1&
$ 0.9118 \pm 0.0014$& 
$3 \pm 2  $&
$5.6 \pm 0.6$ &
9.3 \\

&\ion{Si}{xiv} & 
$1.99 \pm 0.02 $ &
$44^{+20}_{-30}  $&
$2.1 \pm 0.6$ &
3.5&
$ 1.978 \pm 0.012 $ &
$56 \pm 20$&
$2.9 \pm 0.6$ &
4.8 \\

&\ion{Fe}{i} &
$ 6.403 \pm 0.010$ &
$64 \pm 25$&
$2.9 \pm 0.4$  &
7.3&
$ 6.408 \pm 0.007$ &
$82 \pm 10$ &
$4.0 \pm 0.3$&
13  \\

&\ion{Fe}{xxv} (i)&
$6.68  \pm 0.03$ &
20 (fixed) &
$1.0 \pm 0.2$  &
5&
$6.75 \pm 0.02$  &
20 (fixed) &
$0.84 \pm 0.15$&
5.6  \\

&\ion{Fe}{xxvi}  & 
$ 6.99^{+0.02}_{-0.05}$ &
$<39$ &
$1.5 \pm 0.2$   &
7.5&
$ 7.005 \pm 0.012  $ &
 39 (fixed)&
$1.4 \pm 0.2$&
7   \\

\hline

 & & & & & & & & \\
     
      \multicolumn{2}{c}{ } &    
\multicolumn{4}{c}{0.50-0.75 Phase-Interval} &
\multicolumn{4}{c}{0.75-0.95 Phase-Interval} \\

\multirow{8}{*}{HETGS}&\ion{O}{vii} (i)& 
$ 0.5695^{+0.0007}_{-0.0016}$& 
1.1 (fixed)&
$48 \pm 20$ &
2.4&
$ 0.569$ (fixed)& 
1.1 (fixed)&
$<77$&
-- \\

&\ion{O}{viii} & 
$ 0.6525^{+0.0017}_{-0.0012} $& 
$1.3$ (fixed)&
$9 \pm 5$ &
1.8&
$ 0.653$ (fixed)& 
$1.3$ (fixed)&
$<13$&
-- \\

&\ion{Ne}{ix} (i)& 
$ 0.9142 \pm 0.0005$& 
$1.3 \pm 0.5$&
$3.7 \pm 1.2$  &
3.1&
$ 0.9138 ^{+0.0032}_{-0.0015}$ & 
$2.4^{+2.9}_{-1.0}$ &
$3 \pm 2$ &
1.5 \\

&\ion{Ne}{x} &
$ 1.0226^{+0.0002}_{-0.0003} $& 
$<0.7$&
$1.4 \pm 0.5$ &
2.8&
$ 1.0208  \pm 0.0006$ & 
$0.7$ (fixed)&
$1.2 \pm 0.5$ &
2.4\\

&\ion{Mg}{xi} (i)&
$ 1.344 \pm 0.002$ &
$5.1^{+2.0}_{-1.4}$&
$1.2 \pm 0.4$ &
3&
$ 1.343  \pm 0.002$ &
$3 $ (fixed)&
$0.6 \pm 0.3$&
2 \\

 &\ion{Mg}{xii} &

$ 1.470^{+0.003}_{-0.002} $ &
$4 \pm 2$&
$0.7 \pm 0.3$  &
2.3&
$ 1.4697 \pm 0.0015$ &
$3 $ (fixed)&
$0.5 \pm 0.3$&
1.7   \\

&\ion{Si}{xiv} & 
$ 2.003 \pm 0.002$ &
$3.5  \pm 1.3$ &
$0.6 \pm 0.2 $  &
3&
$ 2.002  \pm 0.002 $ &
$3.5$ (fixed)&
$0.5 \pm 0.2 $ &
2.5 \\

&\ion{Fe}{i} &
$ 6.392 \pm 0.005$ &
$17 \pm 5 $ &
$2.7 \pm 0.5$  &
5.4&
$ 6.383^{+0.006}_{-0.003} $ &
$<13 $ &
$1.5 \pm 0.4$  &
3.8\\

&\ion{Fe}{xxvi}  &
$ 6.958 \pm 0.013$ &
$<33$ &
$1.3 ^{+0.7}_{-0.5}$  &
2.6&
$ 6.940 \pm 0.015 $&
$14$ (fixed)&
$0.8 \pm 0.5$&
1.6  \\

& & & & & & & & \\

\multirow{7}{*}{XMM}&
\ion{O}{vii} (i) & 
$ 0.5686 \pm 0.0003 $ &
$1.5 \pm 0.4$ &
$28 \pm 3 $   &
9.3&
$ 0.5685 \pm 0.0006$ & 
$<2.1$ &
$11 \pm 3 $&
3.7 \\

&\ion{O}{viii} & 
$ 0.6521 \pm 0.0011$& 
3 (fixed)&
$5.3 \pm 1.1$ &
4.8&
$ 0.653 \pm 0.002$&
3 (fixed)&
$4.0 \pm 1.0$&
4\\

&\ion{Ne}{ix} (i)& 
$ 0.917 \pm 0.003 $& 
$5^{+3}_{-2}$ &
$3.0^{+0.9}_{-0.5}$ &
6&
0.913 (fixed)&
7 (fixed)&
$1.5 \pm 0.6$&
2.5\\

&\ion{Si}{xiv} & 
$ 1.98 \pm 0.02  $ &
$68 \pm 24$ &
$2.3 \pm 0.7$  &
3.3&
$ 1.97 \pm 0.02$ &
$60$ (fixed)&
$1.2 \pm 0.5 $ &
2.4 \\

&\ion{Fe}{i} &
$ 6.398 \pm 0.008 $ &
$47 \pm 13$ &
$3.3 \pm 0.3$ &
11&
$ 6.40 \pm 0.02$ &
$108 \pm 30 $  &
$2.6 ^{+0.4}_{-0.3}$ &
8.7 \\

&\ion{Fe}{xxv} (i)&
$6.66 \pm 0.03$ &
20 (fixed) &
$0.8 \pm 0.2$  &
4&
$ 6.67 $  (fixed)&
20 (fixed) &
$0.6^{+0.2}_{-0.4} $ &
1.5 \\
 
&\ion{Fe}{xxvi}  & 
$ 6.98 \pm 0.02$ &
$80 \pm 20$  &
$1.7 \pm 0.3 $   &
5.7&
$ 6.97 \pm 0.02$ &
$65 \pm 30$ &
$1.2 \pm 0.3$  &
4  \\


\hline 
 & & &  Id. & E & $\sigma$ & I(10$^{-4}$) & significance& \\
 & & &     & (keV) & (eV) & (ph cm$^{-2}$ s$^{-1}$)  &($\sigma$) & \\
\hline  
  & & & & & & & & \\

\multicolumn{10}{c}{0.95-0.1.04 Phase-Interval} \\

 &  &
\multirow{8}{*}{HETGS}&
\ion{O}{vii} (i)& 
$ 0.569$ (fixed)& 
1.1 (fixed)&
$<140$ & 
--&
\\

 & &
&\ion{O}{viii} & 
$ 0.653$ (fixed)& 
1.3 (fixed)&
$<26$ &
--&
\\

 & & 
&\ion{Ne}{ix} (i)& 
$ 0.921 \pm 0.002$ & 
$2.4$ (fixed)&
$4 \pm 2$  &
2&
\\

 & & 
&\ion{Ne}{x} &
$ 1.020 \pm 0.002$ & 
$2$ (fixed)&
$1.5 \pm 1.2$ &
1.3&
\\

 & & 
&\ion{Mg}{xi} (i)&
$ 1.340 ^{+0.002}_{-0.003}$ &
$3 $ (fixed)&
$1.0 \pm 0.5$ &
2&
\\

 & & 
  &\ion{Mg}{xii} &
$ 1.469 \pm 0.002 $ &
$3 $ (fixed)&
$0.6 \pm 0.4$  &
1.5&
\\

 & & 
&\ion{Si}{xiv} & 
$ 2.005 \pm 0.003$ &
$3.6$ (fixed)&
$0.6 \pm 0.3 $  &
2&
\\

 & & 
&\ion{Fe}{i} &
$ 6.378  \pm 0.011$ &
$12 $ (fixed) &
$1.2 \pm 0.6 $  &
2&
\\

 & & 
&\ion{Fe}{xxvi}  &
$ 6.94 ^{+0.05}_{-0.02} $ &
$25$ (fixed)&
$1.2 \pm 0.9$ &
1.3&
\\
 & & & & & & & & \\

 &  &
\multirow{7}{*}{XMM}&
\ion{O}{vii} (i) & 
$ 0.5694^{+0.0008}_{-0.0026} $  &
$1.5$ (fixed) &
$20 \pm 5 $  &
4&
\\

 &  &
&\ion{O}{viii} & 
$ 0.653$ (fixed)& 
3 (fixed)&
$3 \pm 2$ &
1.5&
\\

 &  &
&\ion{Ne}{ix} (i)& 
$ 0.913 $ (fixed)& 
$7$ (fixed)&
$1.5 \pm 1.1$  &
1.4&
\\

 &  &
&\ion{Si}{xiv} & 
$ 2.003$ (fixed) &
$60$ (fixed) &
$<1.2 $  &
--&
\\

 &  &
&\ion{Fe}{i} &
$ 6.38 \pm 0.03$ &
$<65$&
$1.2 \pm 0.4$ &
3&
\\

 &  &
&\ion{Fe}{xxv} (i)&
$ 6.67$ (fixed)&
20 (fixed) &
$<0.6$  &
--&
\\

 &  &
&\ion{Fe}{xxvi}  & 
$ 6.93 \pm 0.07 $  &
60 (fixed) &
$0.7 \pm 0.4 $ &
1.8&
\\

\hline                                             
\end{tabular}
\end{center}

{\small \sc Note} \footnotesize--- Best-fit values
 of the emission lines  obtained using Gaussian profiles.  
 Uncertainties are  at the  68\% c. l.
 for each shown  parameter.   
\end{table*}


The emission lines are clearly detected during most orbital
  phases, as shown in Table \ref{Line1}.  The HETGS lines are evident
during the phases between 0.04 and 0.75, that is out of the minimum in
the light curve and eclipse.  In spectra 1, 2, and 3 we fixed only the
width of the \ion{O}{vii} line at the values obtained in the averaged
spectrum because of the low statistics below 0.6 keV.  In spectra 2
and 3 we fixed the width of the \ion{O}{viii} line at the best-fit value
obtained in spectrum 1. The other parameters of the \ion{O}{vii} and
\ion{O}{viii} lines were left free to vary in spectra 1, 2, and
3. Furthermore, all parameters associated with the other detected
lines were left free to vary in spectra 1, 2, and 3. In spectrum 4
(phase 0.75-0.95) we fixed both the energies and the widths of the
\ion{O}{vii} and \ion{O}{viii} by obtaining upper limits for the
corresponding fluxes; moreover, we fixed the widths of the \ion{Ne}{x},
\ion{Mg}{xi}, \ion{Mg}{xii}, \ion{Si}{xiv}, and \ion{Fe}{xxvi} lines
at their best-fit values.  In spectrum 5 (phase 0.95-1.04) we found an
upper limit for the \ion{O}{vii} and \ion{O}{viii}, while we fixed the
widths of the other detected lines at their best-fit values.

In the XMM spectra we fixed the width of the weak \ion{Fe}{xxv} line
at 20 eV as in the averaged XMM spectrum.  We fixed the width of the
\ion{O}{viii} line at the corresponding upper limit obtained from
XMM spectrum 1 in spectra from 2 to 5.  In  XMM spectrum 2 we fixed
the width of the \ion{Fe}{xxvi} lines at the corresponding upper limit
obtained in spectrum 1.  In  XMM spectrum 4 we fixed the energy and
the width of the \ion{Ne}{ix} line, the width of the \ion{Si}{xiv}
line, and the width of the \ion{Fe}{xxv} line at their best-fit
values.  In  XMM spectrum 5 the widths were fixed at their best-fit
values.  The equivalent widths in units of eV of the most prominent
lines are shown in Table \ref{Tab_eqw}.

\begin{table*}[ht]
  \caption{Line equivalent widths along the orbital period }
\label{Tab_eqw}      
\begin{center}                                      
\begin{tabular}{l l c c c c c }          
\hline\hline 
 & &
0.04-0.25&
0.25-0.50&
0.50-0.75&
0.75-0.95&
0.95-1.04 \\
\hline       

RGS & \ion{O}{vii} (i)  &
$24 \pm 6$&
$23 \pm 5$&
$25 \pm 4$&
$14 \pm 6$&
$28^{+14}_{-10} $\\

 RGS & \ion{O}{viii}   &
$4 \pm 3$&
$6 \pm 2$&
$6 \pm 2$&
$7 \pm 4$&
$<14 $\\

HETGS & \ion{Ne}{ix} (i-r)  &
$7\pm 5$&
$7\pm 3$&
$5\pm 3$&
$5 \pm 4$ &
$<14$\\

HETGS & \ion{Ne}{x}   &
$4 \pm 2$&
$4 \pm 2$&
$<4$&
$<5$&
$<7$ \\

HETGS & \ion{Mg}{xi} (i)  &
$2.8 \pm 1.1$&
$4.3 \pm 1.4$&
$3 \pm 2$&
$<3$&
$<6$\\

HETGS & \ion{Mg}{xii}   &
$2.6 \pm 1.1$&
$1.8 \pm 1.0$&
$<3$&
$<3$&
$<5$\\

HETGS & \ion{Si}{xiv}   &
$3.3 \pm 1.3$&
$2.3 \pm 1.3$&
$2.2^{+1.4}_{-1.7}$&
$2.1^{+1.4}_{-1.7}$&
$<5$\\

 & & & & & & \\
HETGS & \ion{Fe}{i}   &
$50 \pm 13$&
$46 \pm 12$&
$36 \pm 14$&
$23 \pm 14$&
$<46$\\

EPIC-pn  & \ion{Fe}{i}  &
$54 \pm 11$&
$62 \pm 7$&
$55 \pm 9$&
$50 \pm 13$&
$33 \pm 20$\\

 & & & & & & \\

HETGS & \ion{Fe}{xxvi}   &
$28^{+17}_{-14}$&
$20^{+11}_{-14}$&
$<36$&
$<30$ &
$<67$\\

 EPIC-pn  & \ion{Fe}{xxvi}   &
$31 \pm 10$&
$24 \pm 6$&
$31^{+12}_{-9}$&
$25 \pm 11$&
$<47$\\

\hline\hline 

\end{tabular}
\end{center}                                      

{\small \sc Note} \footnotesize--- The equivalent widths of the most prominent
lines observed with HETGS, RGS, and EPIC-pn. The values are in units of eV,
the errors are at 68\% confidence level.      
 \end{table*}

\section{Discussion}
We analysed one XMM and two HETGS observations of X1822-371, 
studying the averaged
spectrum and the orbital-phase resolved spectra.   The best-fit
  value of the equivalent hydrogen column associated with neutral
  interstellar matter, N$_H \sim 0.16 \times 10^{22}$ cm$^{-2}$, is
  similar to that obtained by \cite{Iaria2001_1822}
  (i.e. $0.11^{+0.3}_{-0.5} \times 10^{22}$ cm$^{-2}$) and to the
  value of $\sim 0.16 \times 10^{22}$ cm$^{-2}$ obtained by
  \cite{Ji_2011}, who  fitted the HETGS spectrum of X1822-371 with a
  power-law component absorbed by local neutral matter.  The values of
  equivalent hydrogen column associated with local neutral matter, 
  N$_{H_{pc}}$, obtained by fitting the XMM and Chandra averaged
  spectra are compatible with each other and consistent with the value
  reported by \cite{Iaria2001_1822} (about $4.4 \times
  10^{22}$ cm$^{-2}$) and by \cite{Ji_2011} (between 5.4 and
  6.3 $ \times 10^{22}$ cm$^{-2}$).  Finally, we found that 
the covered X-ray emitting surface is about $60-65$\%.

The spectra show several emission lines associated with ionised
elements and a fluorescence iron line, as already shown by
\cite{Cottam2001} and \cite{Ji_2011}.

We observed significant residuals in EPIC-pn data below 0.8 keV.  To
remove them we added a black-body component with a temperature of 0.06
keV and a corresponding luminosity of $(6.3 \pm 1.3) \times 10^{34}$
erg/s, a factor of one hundred smaller that the luminosity of the
Comptonised component. We will not discuss this component in more
detail because we cannot distinguish between a real effect of an
instrumental artefact of the Epic-pn below 0.8 keV. Future
observations, e.g. with Astro-H, will allow one to asses the physical
origin of this component.

The 0.1-100 keV extrapolated unabsorbed fluxes of the Comptonised
  component obtained from the XMM and Chandra averaged spectra are $9.2
  \times 10^{-10}$ and $1.3 \times 10^{-9}$ erg cm$^{-2}$ s$^{-1}$,
  respectively.   To check the consistency of the
  measured fluxes in the two observations, we inspected at the long-term
  light curve provided by the All-Sky Monitor (ASM) on board the
  RXTE satellite.  During the  XMM and Chandra observations
  the ASM averaged count rate was $1.3 \pm
  0.2$, and $1.7 \pm 0.2$ c/s, respectively. The ratio between the ASM
  count rate during the Chandra observation, ASM$_{Ch}$, and the XMM
  observation, ASM$_{XMM}$, is $1.3 \pm 0.3$, and the ratio of the 0.1-100
  keV extrapolated unabsorbed Chandra  and XMM flux is 1.4,  compatible 
with the ratio ASM$_{Ch}$$/$ASM$_{XMM}$.
  The proportionality of the ASM count rates to the
  unabsorbed fluxes suggests that the spectral shape of X1822-371 does
  not change significantly with time.
Adopting a distance to the source of 2.5 kpc \citep[][hereafter we
will use this distance]{Mason82}, the extrapolated observed luminosity in the
0.1-100 keV energy range is $7 \times 10^{35}$ erg s$^{-1}$ and
$10^{36}$ erg s$^{-1}$ for the XMM and Chandra averaged spectra,
respectively.

In the following we propose a possible interpretation of the features
observed in the X-ray spectrum of X1822-371. We propose a scenario in
which 1) the modulation of the light curve is mainly caused by the
occultation by the outer accretion disc (section \ref{4d1}); 2) only
1\% of the luminosity produced in the innermost region of X1822-371 is
scattered by an extended optically thin corona into the line of sight
(section \ref{sec1}); 3) the photoionised lines of 
  \ion{O}{viii}, \ion{Ne}{ix}, \ion{Ne}{x}, \ion{Mg}{xi},
\ion{Mg}{xii}, and \ion{Si}{xiv} are produced in a compact region at
the outer radius of the accretion disc (section \ref{4d4}); 4) the
\ion{Fe}{xxvi} line is produced  at a distance from the neutron star less 
than $3.7 \times 10^{10}$ cm (section \ref{4d4}); 5) the \ion{O}{vii} and the
fluorescence iron line are produced in the photoionised surface of the
accretion disc at radii between $2 \times 10^{10}$ and $4 \times
10^{10}$ cm (sections \ref{4d4} and \ref{4d5}); 6) the neutral matter
that partially covers the scattered emission is placed at a distance
from the neutron star comparable to, or larger than, the outer accretion
disc radius (section \ref{4d3}).

\subsection{The presence of an opaque shield}
\label{4d1} 

 Our best-fit model consists of an optically thick Comptonised
  spectrum absorbed by neutral interstellar matter that is partially
  absorbed and Thomson-scattered by local neutral matter.  However,
  even taking into account the Thomson scattering, this model does not
  explain the different values of the unabsorbed flux for the five
  selected phase-intervals.  The 0.4-12 keV unabsorbed flux obtained
  from the XMM data changes along the orbital period from $(7.22
  \pm 0.03) \times 10^{-10} $ erg cm$^{-2}$ s$^{-1}$ in the
  phase-interval 0.25-0.50 down to $(4.05 \pm 0.06) \times 10^{-10} $
  erg cm$^{-2}$ s$^{-1}$ during the eclipse (see Fig. \ref{Fig_cont},
  bottom panel).  \cite{Burderi_2010} gave strong evidence that the
  neutron star in X1822-371 accretes at the Eddington limit while the
  companion star has an overall mass outflow between 3.5 and 7.5 times
  the maximum Eddington accretion rate $\dot{m}_E$ for a 1.4 M$_\odot$
  neutron star. Consequently, we should expect that the unabsorbed flux of the
  system is constant at the Eddington limit along the orbital
  period. The observed changes of the flux can be explained by the
  presence of an opaque shield whose height varies along the orbital
  phase.  The presence of an opaque shield in X1822-371 was already
  discussed by several authors \citep[see e.g.][]{White_holt_1982,
    Mason82, Hellier_mason1989, Bayless2010}, who suggested that the
  outer accretion disc of X1822-371 is geometrically thick and its
  height depends on the azimuthal angle. The 0.4-12 keV unabsorbed
  flux for the phase intervals 0.04-0.25, 0.50-0.75, 0.75-0.95, and
  0.95-1.04 is $83$\%, $91$\%, $76$\%, and $56$\%, respectively, as a
  percentage compared to that during the phase interval 0.25-0.50.
  The highest flux reduction during the phase-interval 0.95-1.04 is
  due to the occultation caused by the companion star.

  On the other hand, the equivalent hydrogen column of the local
  neutral matter and the corresponding covering fraction $f$ are
  correlated with the light-curve modulation (see Figs. \ref{Fig2} and
  \ref{pcfainit}) in the sense that their best-fit values increase
  when the count rate increases and reach the minimum when the count
  rate is minimum (out of the eclipse). This suggests that we cannot
  ascribe the observed light-curve modulation to the local neutral
  matter because if that were the case, we should expect a rise of the
  equivalent hydrogen column of the local matter and/or of the
  covering fraction $f$ when the count rate drops at phases
  0.75--0.95.

  Because we identify the opaque shield with the outer accretion disc,
  we expect that it is the highest when the unabsorbed flux is minimum
  and, conversely, the lowest when the unabsorbed flux is maximum; in
  other words, the height of the outer disc is minimum at phases
  0.25-0.50 and maximum at phases 0.75-0.95.  In agreement with
  \cite{Cottam2001} and other authors we propose that the changes of
  the height at the outer rim is caused by the impact with the outer disc
  of matter coming from the companion star. At the impact point we
  expect the largest height of the outer radius of the accretion disc.

\subsection{The continuum emission and presence 
 of  an optically thin ADC}  \label{sec1}

Although we find the same spectral shape reported by
\cite{Iaria2001_1822}, in this work we will give a different
interpretation of its origin in light of the recent results
reported by \cite{Burderi_2010}, \cite{Bayless2010}, and
\cite{Iaria_2011}, which showed a large derivative of the orbital period
of X1822-371.   \cite{Burderi_2010} showed that the large
  orbital-period derivative of X1822-371 indicates that the system
  accretes at the Eddington limit.  Another indication that the
luminosity of X1822-371 is at the Eddington limit is given by the
ratio $L_X/L_{opt}$, where $L_X$ and $L_{opt}$ are the X-ray and
optical luminosity.  \cite{Hellier_mason1989} showed that the ratio
$L_X/L_{opt}$ for X1822-371 is $\sim 20$, a factor 50 smaller than the
typical value of 1000 for the other LMXBs.  This indicates that the
intrinsic X-ray luminosity is underestimated by at least a factor of
50.  Finally, \cite{Jonker_2001} showed that for a luminosity of
X1822-371 of $10^{36}$ erg s$^{-1}$ the magnetic field intensity of
the neutron star assumes an unlikely value of $8 \times 10^{16}$ G,
while for a luminosity of the source at the Eddington limit it assumes
a more conceivable value of $8 \times 10^{10}$ G.  These three
independent arguments lead us to assume that the intrinsic luminosity
of X1822-371 is close to the Eddington limit; hence the first point
that we discuss is why the observed luminosity of X1822-371 is
$10^{36}$ erg s$^{-1}$ \citep[see e.g.][and this
work]{Hellier_mason1989,Heinz_nowak_2001,parmar2000,Iaria2001_1822}.

Considering that the outer disc is thick and that the
inclination angle of the system is $82.5^\circ \pm 1.5^\circ $, 
 we do not observe the direct emission from the innermost
region and invoke the presence of an extended corona above the disc that
scatters 1\% of the Eddington X-ray luminosity along the line of
sight. A similar scenario was discussed by \cite{McClintock1982} for
4U 2129+47 and by \cite{Hellier_mason1989} for this source. This
scenario is different from that proposed by \cite{Iaria2001_1822} and
other authors, according to which the continuum emission originates in
an extended optically thick corona that has a luminosity of $10^{36}$
erg s$^{-1}$. Indeed, if an extended optically thick corona is
present, a large part of the Eddington luminosity should be
reprocessed there and re-emitted, and the observed luminosity would be almost
$10^{38}$ erg s$^{-1}$, unless one considers the unlikely scenario in
which the emission from the inner region of X1822-371 is extremely
beamed.

The photons produced in the innermost region near the NS are
Comptonised in an optically thick plasma ($\tau \simeq 20$) with a
temperature of 3 keV,  producing the main spectral component
  observed in the X-ray spectrum.  The Comptonising cloud is probably
a compact region near the NS.  X1822-371 has a peculiar geometry for
two reasons: the system is observed at high inclination angle and the
height of the outer accretion disc is large. Combining these two
effects, we infer that we do not observe the direct emission $L_0$ from
the central region, but only the emission scattered by an extended
corona.  The scattered flux, considering $\tau_{C} \ll 1$, is $L_0
\left[(1-e^{-\tau_{C}}) / (4 \pi D^2)\right] \simeq L_0 \tau_{C}/(4
\pi D^2)$, where $D$ is the distance to the source. Therefore the
scattered flux is $\tau_{C}$ times the emission produced near the NS.
Assuming that $\tau_{C} \sim 0.01$ we match the expected luminosity at
the Eddington limit and the observed luminosity of $\sim 10^{36}$
erg/s.

\cite{Jonker_2003} found that the mass function of X1822-371 is $(2.03
\pm 0.03) \times 10^{-2} M_{\odot}$. Assuming a typical NS mass of 1.4
$M_{\odot}$, we derive a companion star mass of 0.408 $M_{\odot}$.
Considering that the orbital period of X1822-371 is 5.57 hours, we
derive that the orbital separation is $a = 1.34 \times 10^{11}$ cm
using the third Kepler law. The Roche lobe of the NS and the companion
star are $R_{L_1} = 9.8 \times 10^{10}$ cm and $R_{L_2} = 3.7 \times
10^{10}$ cm, respectively. For a mass ratio $q=M_2/M_1 \simeq 0.3$,
the tidal truncation radius, that is the maximum possible value of the
accretion disc radius $R_d$, is $ 0.43 a = 5.7 \times 10^{10}$ cm
\citep{Frank,Bayless2010}. We adopt $R_d = 5.7 \times 10^{10}$ cm as
outer accretion disc radius.  The estimated orbital parameters allow
us to infer the size of the optically thin corona, making the
reasonable assumption that the Roche lobe radius $R_{L_2}$ of the
companion star is equal to the radius of the companion star $R_2$.

We can estimate the size of the ADC taking into account the different
count rates observed at mid-eclipse time and out of the eclipse. Since
the spectral shape does not change along the orbit, we impose that the
change in count rates, in and out the eclipse, is mainly due to the
volume of the visible optically thin ADC occulted by the companion
star.  Looking at the HETGS folded light curves of X1822-371 in
Fig. \ref{Fig2}, we observe that the count rate is $C_{ecl} \sim 3$
cts s$^{-1}$ at the mid-eclipse time while it changes between 4 and 6.6
cts s$^{-1}$ out of  the eclipse.  The maximum observed count rate
$C_{out}$ indicates the minimum shielding and allows us to give a
more accurate estimate of the ADC size.  We also note that the values
of $C_{ecl}$ and $C_{out}$ in the EPIC-pn light curves are 37 and 84
cts s$^{-1}$, respectively; we obtain consistent results using EPIC-pn
or HETGS count rates.  Assuming a spherical volume of the ADC, the
scattering volume is proportional to $C_{out}$ out of the eclipse,
while the scattering volume of the ADC subtracted by the volume
occulted by the companion star is proportional to $C_{ecl}$ during the
eclipse. That is
$$
\frac{\frac{4}{3} \pi R_{ADC}^3}{\frac{4}{3}\pi R_{ADC}^3 -\pi R_{2}^2 2 R_{ADC}}
\sim
\frac{C_{out}}{C_{ecl}}\simeq 2.2.
$$
We find that $ R_{ADC} \sim 6 \times 10^{10}$ cm similar to the
accretion disc radius $R_d \sim 5.7 \times 10^{10}$ cm.  In the
following we assume that $R_{ADC} = R_d \sim 5.7 \times 10^{10}$ cm.
Considering the ADC as an extended optically thin cloud with $\tau_{C}
\sim 0.01$, we can estimate the averaged electron density $n_e$ of the
ADC using $\tau_{C} = \sigma_T n_e R_{\rm ADC}$ where $ \sigma_T $ is
the Thomson cross section and $R_{\rm ADC}$ is the radius of the
accretion disc corona; we obtain that the average electron density in
ADC is $n_e \sim 3 \times 10^{11}$ cm$^{-3}$.

The detection of the NS spin period of 0.59 s in X1822-371
\citep{Jonker_2001} also supports a scenario in which an
extended optically thin corona is present.  Indeed, if we assume that
the extended corona is optically thick and the observed Comptonised
spectrum comes from the extended corona, then the extended corona
should have an optical depth of $\tau = 20$, as observed in our fits.
The unscattered photons fraction, $F = F_0 e^{-\tau}$, should be
extremely small because of $\tau \simeq 20$. All photons in the
corona should be scattered with a number of steps to escape of
$N=\max{(\tau,\tau^2)}\simeq \tau^2$ (for $\tau \gg 1$) and their
travelled space should be $N \lambda = \tau^2 \lambda$, where
$\lambda$ is the mean free path of the photons. For an optically thick
ADC with $R_{ADC} =5.7 \times 10^{10}$ cm and $\tau \simeq 20$ we find
$n_e \simeq 5 \times 10^{14}$ cm$^{-3}$, $\lambda \simeq 3 \times
10^{9}$ cm, the average delay of the escaping photons $t_d = \tau^2
\lambda/c \simeq 40$ s, and, finally, the spread in the arrival times
of the coherent photons $\Delta t = \tau \lambda/c \simeq 2$ s.  Note
that $\Delta t$ is almost three times the observed pulse period of
X1822-371 (that is 0.59 s), suggesting that a scenario with an extended
optically thick corona in X1822-371 is not viable because the ADC
should wash out coherent pulsations and prevent the observation of the
0.59 s pulse period.  On the other hand, our scenario predicts an
extended optically thin ADC with an electron density $n_e \sim 3
\times 10^{11}$ cm$^{-3}$ and $\tau_{C} \simeq 0.01$; the mean free
path of the photons in the ADC is $\lambda \sim 5 \times 10^{12}$ cm,
a factor 40 larger than the orbital separation of the system,  in
  agreement with the fact that most photons are not scattered in the
  corona.  We expect that only 1\% of the photons are scattered in
ADC, the escaping photons are scattered once at most and the
average delay of the escaping photons is $ \tau \lambda/c \simeq 2$ s,
similar to the optically thick case. However, because of the absence
of self-occultation in ADC, the spread in the arrival times of
coherent photons depends on the travelled length $l$ in ADC. In the
worst case, for photons scattered in the outer region of the ADC,
$l=R_{ADC}$ then $\Delta t_{max} =R_{ADC}/c \sim 2$ s, while for
photons scattered in inner regions of the ADC, for example at $R=
10^{10}$ cm, we find $\Delta t = 0.3$ s, which allows us to observe the 0.59 s
pulse period, although with a reduced rms amplitude.
 We note that the 1\% scattered emission coming from the innermost
  region is additionally shielded by the outer disc, as discussed in 
the previous
  section. Finally, the different fluxes measured during the XMM and
  Chandra observation can be ascribed to a slightly different optical
  depth of the scattering corona or shielding from the outer disc
  rather than to an intrinsic variation of the source luminosity that
  constantly emits at the Eddington limit (see \cite{Burderi_2010}).

\subsection{Presence of neutral matter around the binary system}
\label{4d3}

We detected the presence of local neutral matter  that
partially covers the X-ray scattering ADC.  We found that the local
neutral matter has an equivalent hydrogen column of around  $5 \times
10^{22}$ cm$^{-2}$ and a covering fraction of  60-65\%.

  We propose that
a large part of the outflowing matter is responsible for the partial
occultation of the extended X-ray scattering ADC, and suggest that the
accreting matter above the Eddington threshold is radiatively expelled
from the system at the inner Lagrangian point. Indeed, as estimated
by \cite{Burderi_2010}, only the 20\% of the mass transferred from the
companion star accretes onto the neutron star of 1.4 M$_{\odot}$.  We
report below our estimation on the equivalent hydrogen column associated
with the local neutral matter considering the mass continuity
equation:
\begin{equation}
(1-\beta) |\dot{M}_2| = 4 \pi r^2 \zeta \rho(r) v(r).
\label{eq1}
\end {equation}
In eq. \ref{eq1}, $\beta$ is the fraction of matter  from 
the companion star that accretes onto the compact object, $\dot{M}_2$ is the 
rate of mass transfer from the companion star, $\zeta$ is a parameter 
that takes into account
a non-spherical  distribution of matter, $\rho(r)$  and $v(r)$
are the density and speed of the matter at a given radius.  
The speed of the matter will be faster than the escape speed $v_{\rm esc}(r)$, 
and hence $v(r) =\eta v_{\rm esc}(r)$ with $\eta$ a parameter larger than 1. 

We  know that the escape speed can be roughly written for a distance 
$r  \gg a$ (with $a$ orbital separation of the binary system) as
\begin{equation}
v_{\rm esc}(r) = \left[2\frac{G (M_1+M_2)}{r}\right]^{1/2},
\label{eq2}
\end {equation}
where $M_1$ and $M_2$ are the masses of the neutron star  and companion star, 
respectively, and G is the gravitational constant.
Combining eq. \ref{eq2} with  eq. \ref{eq1}, we obtain
\begin{equation}
\rho(r)=4.7 \times 10^3 \; (1-\beta) \zeta^{-1} \eta^{-1} \; \dot{m}_{2_E}  \;
(m_1+m_2)^{-1/2} r^{-3/2} \hspace{.2cm} {\rm g \; cm^{-3}}, 
\label{eq3}
\end {equation}
where $\dot{m}_{2_E}$ is the mass transfer rate from the companion star in
 units
of Eddington mass accretion rate and  $m_1$ and $m_2$ are the masses of the 
neutron star and companion star in units of solar masses. 
 We obtain the particle density $n$ from  eq. \ref{eq3} by dividing 
 the density 
$\rho$ by the proton mass and taking into account 
an appropriate fully ionised cosmic mixture of gases ($\mu = 0.615$):
\begin{equation}
n(r) \simeq 4.5 \times 10^{27} \; (1-\beta) \zeta^{-1} \eta^{-1} \; 
\dot{m}_{2_E}  \; (m_1+m_2)^{-1/2} r^{-3/2} \hspace{.2cm} {\rm  cm^{-3}}. 
\label{eq4}
\end {equation}
From the third Kepler law, the orbital separation $a$ 
is 
$$
a = 3.5 \times 10^{10} 
(m_1+m_2)^{1/3} P_h^{2/3}  \hspace{.2cm} {\rm  cm},
$$
where $P_h$ is the orbital period in units of hours. We can rewrite eq. 
\ref{eq4} as
 \begin{equation}
n(r) \simeq 6.9 \times 10^{11} \; (1-\beta) \zeta^{-1} \eta^{-1} \; \dot{m}_{2_E}
  \; (m_1+m_2)^{-1}  P_h^{-1} \left(\frac{r}{a}\right)^{-3/2}  \hspace{-.4cm}
{\rm  cm^{-3}}.
\label{eq5}
\end {equation}

\cite{Burderi_2010} reported a value of  
$\beta \simeq 0.18$ and $ \dot{m}_{2_E} \simeq 5.5$
for a neutron star  mass of 1.4 M$_{\odot}$.
 The orbital period of X1822-371  is $P_h  \simeq 5.57$, therefore  we 
derive
$$
n(a) \simeq 3.1 \times 10^{11} (\zeta \eta)^{-1}  \hspace{.2cm} 
{\rm  cm^{-3}}.
$$ 

Supposing a constant particle density along the line of sight,
  we can determine the
 equivalent hydrogen column 
$N$ associated with the neutral matter  using  $N= n(a) \times a$, where
$a \simeq 1.3 \times 10^{11}$ cm for X1822-371, that is    
$$
N \simeq 4 \times 10^{22} (\zeta \eta)^{-1}  \hspace{.2cm} 
{\rm  cm^{-2}}.
$$ 
This value is similar to the equivalent hydrogen column of the cold
matter that partially occults the central region ($\simeq 5 \times
10^{22}$ cm$^{-2}$) under the hypothesis that the product $\zeta \eta
$ is close the unity. This agrees with our scenario, as
$\zeta$ takes into account the non-spherical distribution of matter
and assumes values lower than 1, while $\eta$ is a parameter that
indicates the speed of the matter in units of the escape
speed, the lowest possible  speed under the hypothesis that
the matter leaves the binary system.  The presence of matter in the
outer region of the disc or in a region surrounding the entire system
was also proposed by \cite{harla1997} and \cite{cowley_2003}, observing
that the H$\alpha$ line is formed in a region not strongly eclipsed and
that the H$\alpha$ velocity variation is very small. These results were 
confirmed recently by \cite{peris2012}

\subsection{The ionised matter}
\label{4d4}

We detected ionised emission lines associated with
He-like and H-like ions that are \ion{O}{vii}, \ion{O}{viii},
\ion{Ne}{ix}, \ion{Ne}{x}, \ion{Mg}{xi}, \ion{Mg}{xii},
\ion{Si}{xiii}, \ion{Si}{xiv}, and \ion{Fe}{xxvi}. The \ion{Ne}{ix}
line is identified with the intercombination line during all 
orbital phases except for that during the eclipse (phase interval
0.95-1.04) when it is identified with the resonance line.   We
  found that all the observed emission lines are clearly detected
  during the orbital phases between 0.04 and 0.75. At the
  phase interval 0.75-0.95 we  clearly detected  the
  \ion{O}{vii} and \ion{O}{viii} in the RGS spectrum, of the
  neutral/partially-ionised iron line in the HETGS and EPIC-pn spectra,
  and, finally, of the \ion{Fe}{xxvi} line in the Epic-pn spectrum;
  the other lines have a significance lower than 3$\sigma$ that does
  not allow us to state with certainty whether they are present at those orbital
  phases even though the small associated errors with the line
  energies (which are all compatible with the expected rest-frame energies)
  suggest their presence. At the eclipse (phase-interval
  0.95-1.04) only the \ion{O}{vii} line and the
  neutral/partially-ionised iron line are clearly identified in the
  RGS and EPIC-pn spectrum, respectively.

The same emission lines were already observed by \cite{Cottam2001},
who  analysed a
shorter Chandra observation. However, because of the poorer statistics,
the lines were observed only during the phases 0-0.5.
\cite{Cottam2001} suggested that the lines originate from the inner
face of the bulge at the outer radius of the accretion disc, where the
stream of accreting matter impacts the outer region of the disc.
During the phases 0-0.5 we directly observed  the inner face of the
bulge, whereas in the phase-interval 0.5-1 the lines are not observed
because the bulge is facing the opposite direction with respect to the
observer.  

We report the equivalent widths of the most prominent lines at
different phases in Table \ref{Tab_eqw}.  The equivalent widths of
  the emission lines associated with \ion{O}{viii}, \ion{Ne}{ix},
  \ion{Ne}{x}, \ion{Mg}{xi}, \ion{Mg}{xii}, and \ion{Si}{xiv} do not
  show significant changes at the phase intervals where they are
  clearly detected, implying that the line and continuum fluxes
change in the same way.  The equivalent width of the \ion{O}{vii} 
  seems to show a reduction at phase 0.75-0.95, and the equivalent
  width of the \ion{Fe}{xxvi} line does not vary along the orbit,
  although we have no  clear detection of this line during the
  eclipse and more investigations at this phase interval are
  needed; however, this line cannot originate in the  same region
  of the bulge because it has largest associated ionisation parameter
with respect the other lines.

The He-like triplets of \ion{O}{vii}, \ion{Ne}{ix}, and \ion{Mg}{xi}
show that the intercombination lines are dominant with respect to the
forbidden and resonance lines at all  phases.  In the following we
report our plasma diagnostics using the best-fitting values obtained
from the averaged spectrum, because they provide the best constraints.
\begin{figure*}
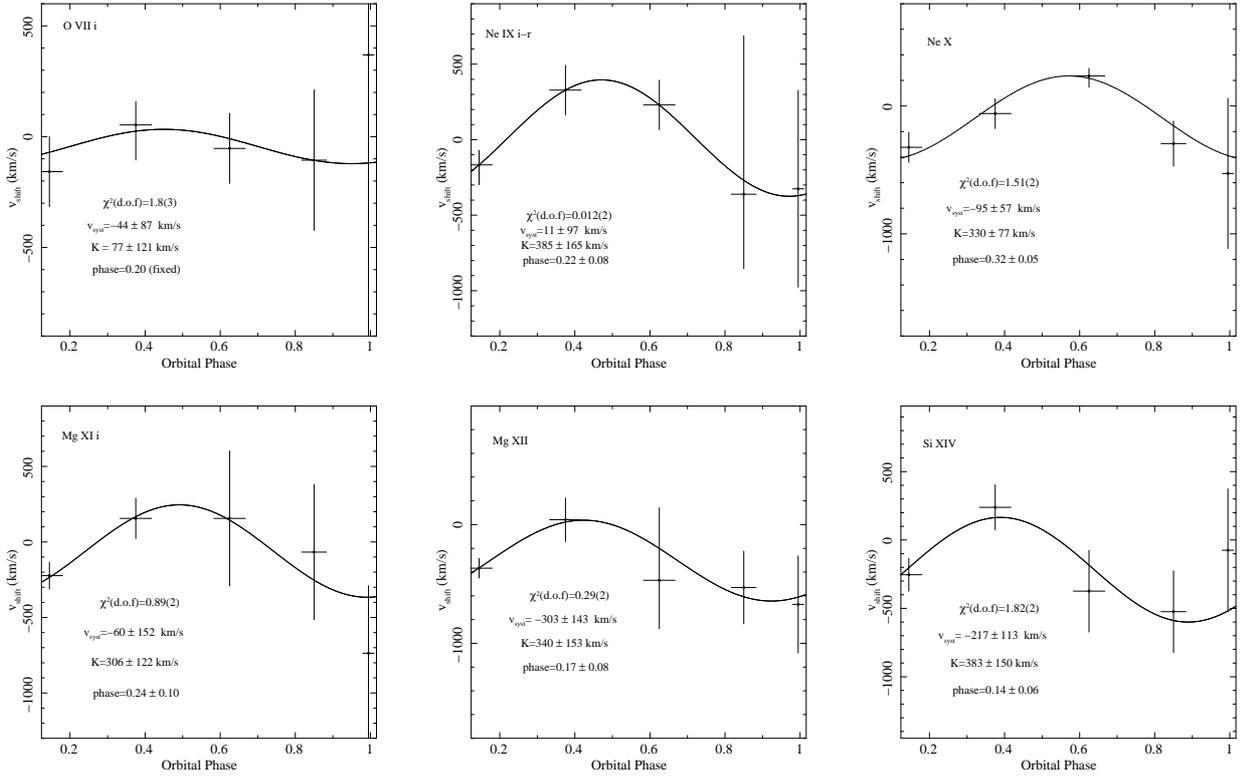

\includegraphics[height=5.3cm, angle=0]{fig10a.ps}
\includegraphics[height=5.3cm, angle=0]{fig10b.ps}
\includegraphics[height=5.3cm, angle=0]{fig10c.ps}\\
\includegraphics[height=5.3cm, angle=0]{fig10d.ps}
\includegraphics[height=5.3cm, angle=0]{fig10e.ps}
\includegraphics[height=5.3cm, angle=0]{fig10f.ps}
\caption[]{ Best-fit parameters of the Doppler modulation
  associated with the \ion{O}{vii}, \ion{Ne}{ix}, \ion{Ne}{x},
  \ion{Mg}{xi}, \ion{Mg}{xii}, and \ion{Si}{xiv} emission lines
  adopting a sinusoidal function as discussed in the text.
  Uncertainties are at 68\% c. l.  }
\label{doppler}
\end{figure*}
From the line intensities of the \ion{O}{vii} triplet, we infer that
the origin of the lines is a photoionised plasma with an electron
density $n_e$ higher than $10^{12}$ cm$^{-3}$ \citep[see the bottom
right panel in Fig. 11 in][]{porquet_dubau00}.  Our analysis indicates
that the $R$ parameter (ratio of the forbidden over intercombination
line intensity) is $R <0.2$ for \ion{O}{vii}, \ion{Ne}{ix} and
\ion{Mg}{xi}, suggesting a plasma density $>10^{12}$ cm$^{-3}$ for
\ion{O}{vii}, $>10^{13}$ cm$^{-3}$ for \ion{Ne}{ix}, and $>10^{14}$
cm$^{-3}$ for \ion{Mg}{xi} \citep[see][]{porquet_dubau00}.  Assuming
that the \ion{Ne}{ix} and \ion{Mg}{xi} are produced in the bulge, we
derive that the bulge has an electron density $n_e>10^{14}$ cm$^{-3}$.

\cite{kallman_bautista_01} studied the properties of a photoionised
plasma at high density ($10^{16}$ cm$^{-3}$) and derived that the
ionisation parameter $\xi$ is between 100 and 400 for \ion{Ne}{ix},
\ion{Ne}{x}, \ion{Mg}{xi}, \ion{Mg}{xii}, and \ion{Mg}{xii} and
\ion{Si}{xiv}, respectively, and $\xi \simeq 1000$ for \ion{Fe}{xxvi}.
We can estimate the electron density of the photoionised plasma using
the relation $\xi =L_X/(n_e r^2)$ \citep[see][]{krolik1981}, where
$L_X$ is the X-ray luminosity of the source, $\xi$ the ionisation
parameter, $n_e$ the electron density of the matter, and $r$ the
distance of the photoionised matter from the central source. Adopting
an X-ray luminosity of $ 2 \times 10^{38}$ erg/s, and placing the
bulge at $6 \times 10^{10}$ cm (the outer radius of the accretion
disc), we derive an electron density $n_e \simeq 5 \times 10^{14}$
cm$^{-3}$ that is compatible with the value of $n_e>10^{14}$
cm$^{-3}$ estimated above from the study of the triplets.

To estimate the temperature of the emitting plasma we use the $G$
parameter inferred in section \ref{trip}, which is $4.3^{+2.9}_{-1.5} $,
$4.4^{+3.9}_{-1.3}$, and $3.7^{+3.7}_{-1.5}$ for \ion{O}{vii},
\ion{Ne}{ix} and \ion{Mg}{xi}, respectively. According to the
calculations outlined in \cite{porquet_dubau00}, we determine that the
temperature of the plasma should be $<5 \times 10^5$ K, $<8 \times
10^5$ K, and $ <10^6$ K for \ion{O}{vii}, \ion{Ne}{ix} and
\ion{Mg}{xi}, respectively.  \cite{Cottam2001} detected a RRC feature
associated with \ion{Ne}{ix} with a temperature of $13 \pm 7 $ eV
corresponding to a plasma temperature of $(1.5 \pm 0.8) \times 10^5$
K. We looked for a similar feature in our HETGS observation and found
only an upper limit to the plasma temperature of 14 eV corresponding
to $1.6 \times 10^5$ K.  We conclude that the region where the
\ion{Ne}{ix}, \ion{Ne}{x}, \ion{Mg}{xi}, \ion{Mg}{xii}, and
\ion{Mg}{xii} and \ion{Si}{xiv} lines originate has an electron
density of $n_e \simeq 5 \times 10^{14}$ cm$^{-3}$ and a temperature
of $\sim 1.6 \times 10^5$ K.

Recently, \cite{Ji_2011}, analysing the same Chandra data sets
  as used for this work, measured a Doppler-shift of several lines.
  \cite{Ji_2011} divided the folded light curve into 40 phase-intervals
  (separated by 0.025 in phase) and adopted a phase window of 0.16
  width in phase, sliding along the 40 intervals from which they
  extracted the spectra.  In this paper we follow a different
  approach, obtaining information about the lines from the five
  independent phase-intervals discussed above.  Below we present the
  Doppler-shift velocities of the \ion{O}{vii} line observed with the
  RGS instrument, and of the \ion{Ne}{ix}, \ion{Ne}{x}, \ion{Mg}{xi},
  \ion{Mg}{xii}, and \ion{Si}{xiv} lines observed with the HETGS
  instruments.
\begin{table*}[ht]
  \caption{Best-fit parameters of the Doppler-shift modulation of the most
 prominent emission lines.}
\label{doppler}      
\begin{center}                                      
\begin{tabular}{l c c c c c c}          
\hline\hline 
Line  & 
Rest-frame energy &  
$v_{syst}$  &  
$K$ & 
$\phi_0$ & 
$\chi^2(d.o.f)$& 
Prob. chance. improv. \\

     &
(keV)  &
km/s &
km/s & & & (\%) \\

\hline   
\ion{O}{vii} &
0.5687 &
$-44 \pm 87$ &
$ 77 \pm 121$ &
0.20 (fixed) &
1.8(3)&
52.6\\

\ion{Ne}{ix} (i-r)&
0.9149(i); 0.9220(r) &
$-11 \pm 97$ &
$ 385 \pm 165$ &
$0.22 \pm 0.08$  &
0.012(2)&
99.9\\

\ion{Ne}{x} &
1.0218 &
$-95 \pm 57$ &
$ 330 \pm 77$ &
$0.32 \pm 0.05$  &
1.51(2)&
94.1\\

\ion{Mg}{xi}  (i)&
1.3433 &
$-60 \pm 152$ &
$ 306 \pm 122$ &
$0.24 \pm 0.10$  &
0.89(2)&
88.7\\

\ion{Mg}{xii}&
1.4723 &
$-303 \pm 143$ &
$ 340 \pm 153$ &
$0.17 \pm 0.08$  &
0.29(2)&
94.7\\

\ion{Si}{xiv}&
2.0055 &
$-217 \pm 113$ &
$ 383 \pm 150$ &
$0.14 \pm 0.06$  &
1.82(2)&
78.5\\

\ion{Fe}{i}&
6.400 &
$-396 \pm 104$ &
$ 323 \pm 155$ &
$0.07 \pm 0.07$  &
1.74(2)&
71.9\\

\ion{Fe}{xxvi}&
6.966 &
$-608 \pm 210$ &
$ 235 \pm 290$ &
$0.83$ (fixed)  &
2.07(3)&
60.0\\

\hline\hline 

\end{tabular}
\end{center}                                      

{\small \sc Note} \footnotesize---  The parameters $v_{syst}$, $K$, 
and $\phi_0$  are defined in the text. The probability of chance 
improvement is obtained with respect to a constant function. The errors 
are at 68\% c.l.
 \end{table*}
\begin{figure*}
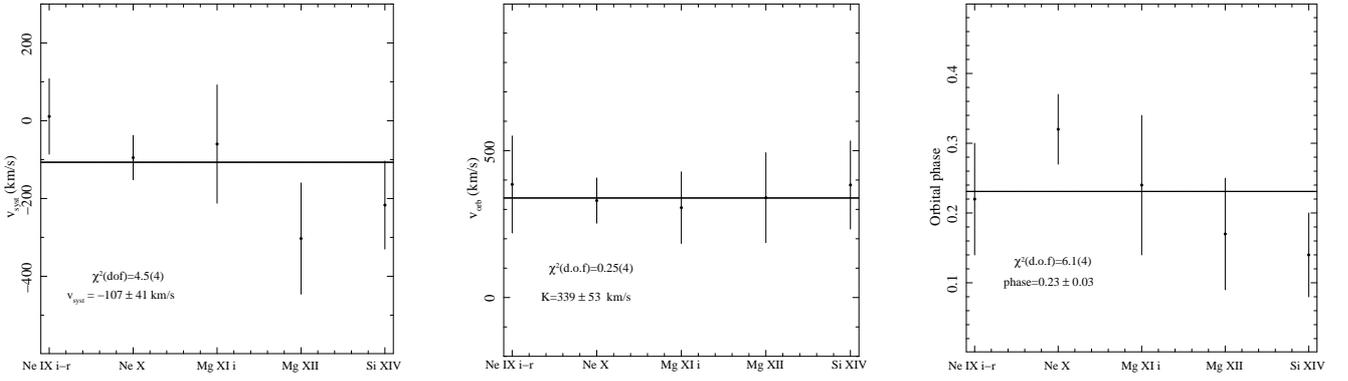

\includegraphics[height=5.4cm, angle=0]{fig11a.ps}
\includegraphics[height=5.4cm, angle=0]{fig11b.ps}
\includegraphics[height=5.4cm, angle=0]{fig11c.ps}
\caption[]{ Best-fit values of the parameters $v_{syst}$ (left panel),
  $K$ (middle), and $\phi_0$ (right panel) for the \ion{Ne}{ix},
  \ion{Ne}{x}, \ion{Mg}{xi}, \ion{Mg}{xii}, and \ion{Si}{xiv}
  lines. Uncertainties are at 68\% c. l. }
\label{doppler_res}
\end{figure*}
 We estimate the Doppler-shift velocities of the \ion{O}{vii},
\ion{Ne}{ix}, \ion{Ne}{x}, \ion{Mg}{xi}, \ion{Mg}{xii}, and
\ion{Si}{xiv} lines for each phase-interval using the best-fit values
of the line energies obtained from RGS data for the \ion{O}{vii} line
and from HETGS data for the other lines, using the best-fit values of the
line energies shown in Table \ref{Line1}.  This analysis was not possible
for the \ion{O}{viii} line because of larger uncertainties.

To estimate the Doppler-shift modulations we fitted the Doppler-shift
velocities, $v_{shift}$, obtained for each phase-interval using the relation
\begin{equation}
\label{eqsin}
v_{shift} = v_{syst}+K \sin (2 \pi (\phi-\phi_0)/P),
\end{equation}
 where the free parameters are $ v_{syst}$, the systemic velocity in
units of km/s, $ K$, the semi-amplitude velocity in units of km/s, and
$\phi_0$, the orbital phase at which the sinusoidal function is
null. The orbital period $P$ in units of phase is kept fixed to 1.
The best-fit parameters for each line are shown in Table \ref{doppler}.

 Initially, we focus our attention on the sinusoidal modulation of
  the \ion{Ne}{ix}, \ion{Ne}{x}, \ion{Mg}{xi}, \ion{Mg}{xii}, and
  \ion{Si}{xiv} lines; the best-fit values of $v_{syst}$, $K$,
  and $\phi_0$ are similar for all these lines, suggesting that these
  five lines are probably produced in the same region. We plot the
  obtained best-fit parameters of the sinusoidal modulation in Fig.
  \ref{doppler_res} (left, middle, and right panel, respectively).
  Under the assumption that the emitting region is the same for the
  five lines, we fit their values of $v_{syst}$, $K$, and
  $\phi_0$ using a constant and finding self-consistent results.  We
  obtain in this way the following average values: $v_{syst} = -107
  \pm 41$ km/s and $\chi^2(d.o.f.)=4.5(4)$, $K = 339 \pm 53$
  km/s and $\chi^2(d.o.f.)=0.25(4)$, and, finally, $\phi_0 = 0.23 \pm
  0.03$ and $\chi^2(d.o.f.)=6.1(4)$.

The systemic velocity, $v_{syst}$, is similar for the five lines with
a value of $v_{syst} = -107 \pm 41$ km/s (see Fig. \ref{doppler_res},
left panel), which is compatible with the systemic velocity of -106
km/s and -101 km/s obtained by \cite{cowley_2003} for the \ion{He}{ii}
$\lambda$4686 and H$\alpha$ lines observed in X1822-371.

The values of $\phi_0$ obtained for the five lines are similar, and
fitting them with a constant, we obtain $\phi_0=0.23 \pm 0.03$ and a
$\chi_{red}^2(d.o.f.)=1.5(4)$, which is acceptable.  We find that the
emitting region has null radial velocity at phases $0.23 \pm 0.03$ and
$0.73 \pm 0.03$ which is when the radial velocity changes from negative
to positive and from positive to negative values, respectively. The
detection of a Doppler modulation implies that the \ion{Ne}{ix},
\ion{Ne}{x}, \ion{Mg}{xi}, \ion{Mg}{xii}, and \ion{Si}{xiv} lines are
produced in a compact region in the bulge at
  the outer accretion disc, as
suggested by \cite{Cottam2001}.  Moreover, the phases between 0.6 and
0.9 correspond to the minimum of the modulation out of the eclipse in
the light curve (see Fig. \ref{Fig2}), which is when the outer
accretion disc is the tallest and, consequently, when the scattered
emission is mostly occulted by the outer disc.  Our results suggest
that the outer region of the disc with the largest height and the
bulge is the same region as the one that covers a phase interval longer than
  0.3; while the  region where the
  \ion{Ne}{ix}, \ion{Ne}{x}, \ion{Mg}{xi}, \ion{Mg}{xii}, and
  \ion{Si}{xiv} lines are produced might be 
  (a  part of) the bulge illuminated by the central source.

\begin{figure*}
\includegraphics[height=7.4cm, angle=0]{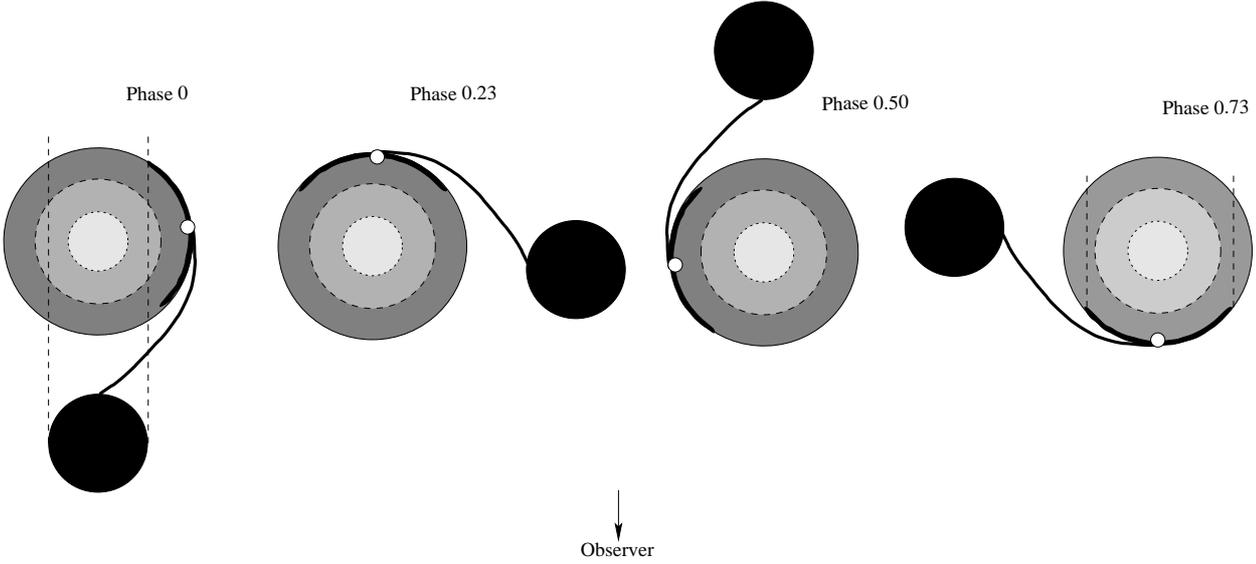}
\caption[]{  Illustration of the geometry of X1822-37.  The
  figure is to scale except for the size of the region in the bulge where
  the \ion{O}{viii}, \ion{Ne}{ix}, \ion{Ne}{x}, \ion{Mg}{xi},
  \ion{Mg}{xii}, and \ion{Si}{xiv} lines originate (white circle). The
  black-coloured arc at the outer radius of the disc indicates where
  the outer disc is the highest, which is the location of the bulge.
  The dashed and dotted circumferences on the disc indicate where the
  \ion{O}{vii} and \ion{Fe}{i}-\ion{Fe}{xv} lines are produced,
  respectively.  }
\label{fig_phase}
\end{figure*}

We show the inferred geometry of X1822-371 in
  Fig. \ref{fig_phase}; the region in the bulge where the lines
  originate (indicated with a white circle) is not to scale because
  its size is uncertain.  Adopting the estimate of the solid angle
subtended by this region inferred by \cite{Cottam2001} and the outer
radius of the disc ($6 \times 10^{10}$ cm) estimated above, we find an
illuminated surface area of $\sim 8.3 \times 10^{20}$ cm$^2$.
Assuming, arbitrarily, that the height $h$ and the subtended arc $l$
of the illuminated region have the same size, we find $h=l \sim 3
\times 10^{10}$ cm; the subtended angle  in this case is
$\sim28^\circ\!$ corresponding to a phase interval of 0.07 that is
shorter than that subtended by the whole bulge, which is 0.3. On the
other hand, assuming that the whole inner face of the bulge is
illuminated by the central source, we estimate that the vertical scale
height is $h \sim 7.3 \times 10^{9}$ cm.  Using detailed modelling of
the X-ray, UV, optical, and IR light curves, \cite{White_holt_1982},
\cite{Mason82}, and \cite{Hellier_mason1989} all inferred a
phase-dependent vertical structure along the outer radius of the disc
that is dominated by the bulge with a maximum vertical height between
0.6 and 1.6 $\times 10^{10}$ cm depending on the model. This is
consistent with the vertical height that we estimate assuming that the
whole inner face of the bulge emits the observed lines, while a
discrepancy of a factor 2 is obtained for the first case ($h \simeq
l$).

At phase 0.73 the outer part of the
  illuminated compact region is facing  the observer and changes
  its radial velocity from positive to negative values. At the same
  phase the outer region of the disc is the tallest and occults a
  large part of the direct emission coming from the inner disc
  surface. The black-coloured arc at the outer radius of the disc
  shown in Fig. \ref{fig_phase} indicates where the outer disc is the
  tallest, that is the whole size of the bulge.

The semi-amplitude velocities are very similar for the five lines, we
find $K = 339 \pm 53$ km/s (see Fig. \ref{doppler_res}, middle panel)
with a $\chi^2(d.o.f.)=0.25(4)$.  The maximum of the blueshift
velocities ($\sim 360$ km/s) is observed at phases between 0.45 and
0.51 and is compatible with the value inferred by \cite{Cottam2001},
who deduced a blueshift velocity of $\sim 360$ km/s at phase 0.50 and
suggested that we are observing the line emissions from the bulge that
was expected to form in the shock-heated, colliding material
\citep[see e.g.][]{livio_1986}.  Our results suggest that the lines
associated with the \ion{Ne}{ix}, \ion{Ne}{x}, \ion{Mg}{xi},
\ion{Mg}{xii}, and \ion{Si}{xiv} transitions are produced in the bulge
illuminated by the central source.

The \ion{O}{vii} line does not show a significant sinusoidal
modulation. We obtain that $ K = 77 \pm 121$ km/s and the probability
of chance improvement with respect to a constant is only 52.6\%.  The
flux and the equivalent width of this line drop during the phase
interval 0.75-0.95 (see Tables \ref{Line1} and \ref{Tab_eqw}). These
arguments suggest that the \ion{O}{vii} line is produced in a
different region.  Because the equivalent width of the \ion{O}{vii}
line drops at phases between 0.75 and 0.95, we argue that at those
phases its emitting region is occulted by the bulge and the outer
disc, implying that the \ion{O}{vii} line is produced in an inner
region with respect to the other ionised lines.  We can speculate that
the \ion{O}{vii} line is produced on the accretion disc surface
illuminated by the central source.  The half width at half maximum
(HWHM) of the line \ion{O}{vii} corresponds to a velocity of about
$680 \pm 130$ km/s, and, consequently, to an accretion disc radius of
$(4.0^{+2.0}_{-1.1}) \times 10^{10}$ cm.  We estimate the electron
density of the photoionised plasma assuming a luminosity at the
Eddington limit ($2 \times 10^{38}$ erg s$^{-1}$) using the relation
$\xi =L_X/(n_e r^2)$ and obtain a plasma density of about $1.3 \times
10^{16}$ cm$^{-3}$, which agrees with our estimation from the study of
the \ion{O}{vii} triplet that $n_e$ is larger than $10^{12}$
cm$^{-3}$. The radius where the \ion{O}{vii} line originates is
comparable with the companion star radius (that is $3.7 \times
10^{10}$ cm), which explains why the equivalent width of the
\ion{O}{vii} line drops at phases between 0.75 and 0.95 and increases
again during the eclipse.  Indeed, at phases 0.75-0.95, the bulge is
interposed between the central source and the observer and its large
azimuthal extension covers the region of the disc that emits the line,
while during the eclipse this region is only marginally occulted by
the companion star.  The \ion{O}{vii} line is clearly detected in the
RGS spectrum along the whole orbital period. Because we observe this
line at phase interval 0.75-0.95, it is probable that we are observing
both direct and scattered line photons of \ion{O}{vii}. Indeed, if we
had observed only the direct line photons, we should not have detected
the \ion{O}{vii} line at this phase interval.

Furthermore, even if the emission lines associated with the
\ion{Ne}{ix}, \ion{Ne}{x}, \ion{Mg}{xi}, \ion{Mg}{xii}, and
\ion{Si}{xiv} transitions are not clearly detected in the phase
interval 0.75-0.95, we tentatively suggest that we are observing both
direct and scattered emission from the illuminated region of the
bulge; this scenario seems reasonable considering that the optically
thin extended corona extends to the outer disc, implying that even
when the outer face of the bulge is orientated towards the observer,
we expect that part of the emission from the illuminated region is
scattered into the line of sight by the extended optically thin
corona.  However, more investigations with observations with larger
statistics are needed to confirm this scenario.

We show in Fig. \ref{fig_phase} the geometry of X1822-371 at phase 0
(during the eclipse) and at phase 0.73 (when the shielding by the
bulge is maximum); during the eclipse the companion star occults the
region of the disc where the \ion{O}{vii} line was produced (indicated
with a dashed circumference in Fig. \ref{fig_phase}) less than the
bulge occults the same region at phase 0.73.

 We were unable to estimate the Doppler motion associated with the
  \ion{O}{viii} emission line because of the large errors associated
  with its centroid energy. However, we suggest that this line is produced in
  the illuminated face of the bulge for two reasons: 1) its equivalent
  width does not change along the orbital phase (see
  Tab. \ref{Tab_eqw}) similarly to the equivalent width of the
  \ion{Ne}{ix}, \ion{Ne}{x}, \ion{Mg}{xi}, \ion{Mg}{xii}, and
  \ion{Si}{xiv} emission lines; 2) the ionisation parameter of the
  \ion{O}{viii} is near 100 \citep[see][]{kallman_bautista_01},
  similar to that obtained for the lines produced in the bulge
  (i.e. between 100 and 400). 

The \ion{Fe}{xxvi} emission line does not show  Doppler
modulation. Initially, we fitted its Doppler-shift velocities with a
constant and then with a sinusoidal function, obtaining a probability
of chance improvement of only 60\%. The corresponding equivalent width
is constant at the phases 0.04--0.95 and, during the eclipse, it shows
an upper limit compatible with the values obtained at the other phases
(see Table \ref{Tab_eqw}).  We have only a marginal detection of  the
 \ion{Fe}{xxvi} emission line during the eclipse. In
  agreement with \cite{Ji_2011}, we argue that this line may be
  produced at $r<3.7 \times 10^{10}$ cm (i.e. the size of the Roche
  lobe radius of the companion star).

\subsection{The neutral/partially ionised iron line}
\label{4d5}

We find an emission line at 6.4 keV in HETGS and Epic-pn spectra,
consistent with fluorescent emission from \ion{Fe}{i}-\ion{Fe}{xv}
\citep{kallman_2004} with an ionisation parameter $\xi \le 10$, similar
to the one associated with \ion{O}{vii}.   The broadening of the
  line at 6.4 keV is $18 \pm 3$ eV as measured by the HETGS (see Table
  \ref{Averaged_line}). The width obtained from the HETGS data is
  compatible with the value inferred by \cite{Ji_2011}.

We do not  significantly detect  Doppler modulation of this
emission line (see Table \ref{Tab_eqw}).  Its flux is minimum at
phases between 0.75-1.04, during the occultation by the bulge and the
companion star.  This can be explained by  assuming that the fluorescence
 iron line is produced at the surface of the accretion disc
similarly to the \ion{O}{vii} line, taking also into account that the
ionisation parameter is comparable for the two lines.  Using the HETGS
best-fitting parameters and assuming that the broadening of the
neutral iron line is caused by a Keplerian motion of the plasma, we find
that it should originate at a distance from the neutron star of
$(1.9^{+0.8}_{-0.5}) \times 10^{10}$ cm, that is at a radius a factor
of two smaller than that at which the \ion{O}{vii} line is produced.
Since the radius of the companion star is $3.7 \times 10^{10}$ cm, 
the emitting region of the iron line is occulted 
both  during the eclipse (phase 0) and during the occultation by the opaque
shield (phase 0.73). We show in Fig. \ref{fig_phase} the radius where 
the fluorescence iron line is produced with a dotted circumference.

\section{Conclusions}

We analysed one XMM-Newton  observation and two Chandra 
 observations  of X1822-371.  We found  that the continuum
emission is well-fitted using a Comptonised component partially absorbed 
by local neutral matter.  In the 
following we summarise our results:

\begin{itemize}
 
\item We inferred the presence of an extended optically thin corona.  The
  Comptonised component, with electron temperature of 3 keV and
  optical depth of 20, is produced in a compact region near the
  neutron star.  In line with the work of \cite{Burderi_2010}, we
  assume that the intrinsic luminosity is $2 \times 10^{38}$ erg/s
  (the Eddington luminosity for a neutron star of 1.4 M$_{\odot}$)
  although the observed luminosity is $10^{36}$ erg/s.  This
  discrepancy is caused by the high system inclination and by the
  geometrically thick outer accretion disc that occults the innermost
  region. We observed only the emission that is scattered into the line
  of sight by an extended optically thin ($\tau \simeq 0.01$) corona.

\item The partial covering component is due to neutral matter located
  at a radius larger than the outer disc radius.  The fraction of the
  covered emitting surface is 60\% along the whole orbit and the
  equivalent hydrogen column is between 4.5 and $5 \times 10^{22}$
  cm$^{-2}$. Since the mass transfer in X1822-371 is not conservative
  \citep{Burderi_2010,Bayless2010}, we suggest that the local neutral
  matter could be the matter outflowing from the companion star, which
  is swept away by the radiation pressure  in the proposed
    scenario.

\item We observed the emission lines associated with \ion{Ne}{ix},
  \ion{Ne}{x}, \ion{Mg}{xi}, \ion{Mg}{xii}, \ion{Si}{xiii}, and
  \ion{Si}{xiv}. These lines show Doppler modulation, suggesting that
  they are produced in a compact region.   In agreement with
    previous authors, we found that  these lines are produced in the
  inner  illuminated face of the bulge at the outer radius of the
  accretion disc distant $6 \times 10^{10}$ cm from the NS.  The
  density of the photoionised emitting plasma is $\sim 5 \times
  10^{14}$  cm$^{-3}$.  We detected the \ion{O}{viii} emission line
    in the RGS spectrum at all phase intervals with a marginal detection
 during
    eclipse. The large  errors on its centroid energy do not
    allow us to find a clear Doppler modulation of this line. However,
    the behaviour of its equivalent width, similar to those of the
    \ion{Ne}{ix}, \ion{Ne}{x}, \ion{Mg}{xi}, \ion{Mg}{xii},
    \ion{Si}{xiii}, and \ion{Si}{xiv} lines, and its similar ionisation
    parameter suggest that the \ion{O}{viii} emission line is produced
    in the illuminated bulge as well.

\item  The \ion{Fe}{xxvi} line was observed at all phases with a
    marginal detection during the eclipse.  This line has a large
    ionisation parameter and does not show significant Doppler modulation. In
    agreement with \cite{Ji_2011}, we suggest that this line is
    produced at a distance from the compact object less than $3.7
    \times 10^{10}$ cm.

\item We observed the \ion{O}{vii} and a prominent fluorescence iron
  line that can be associated with \ion{Fe}{i}-\ion{Fe}{xv}.  These
  lines do not show   significant Doppler-modulation and are
  produced in the photoionised surface of the accretion disc, at
  $4.0^{+2.0}_{-1.1} \times 10^{10}$ cm and $1.9^{+0.8}_{-0.5} \times
  10^{10}$ cm, respectively.  The \ion{O}{vii} emission line is
  partially occulted only at phases between 0.75--0.95 by the bulge
  while the fluorescence iron line is occulted also during the eclipse
  considering that the radius of the companion star is $3.7 \times
  10^{10}$ cm.

\end{itemize}

\begin{acknowledgements} 
  We are grateful to the anonymous referee for his/her useful
  suggestions.  LB and TD acknowledge support from the European
  Community's Seventh Framework Program (FP7/2007-2013) under grant
  agreement number ITN 215212 Black Hole Universe.
\end{acknowledgements} 
\bibliographystyle{aa} 
\bibliography{citations}
\end{document}